\documentclass[aps,prb,twocolumn,superscriptaddress,floatfix,longbibliography,nofootinbib]{revtex4-2}

\usepackage[utf8]{inputenc}
\usepackage[nice]{nicefrac}
\usepackage{graphicx,color}
\usepackage{amsmath,amssymb,bm,bbm}
\usepackage{amsmath}
\usepackage{braket}
\usepackage{xcolor}
\usepackage{float}

\usepackage[nice]{nicefrac}

\usepackage{physics} 

\usepackage[T1]{fontenc}
\usepackage[utf8]{inputenc}

\DeclareUnicodeCharacter{2212}{~}

\usepackage[plainpages=false,pdfpagelabels,colorlinks=true,linkcolor=red,urlcolor=blue,citecolor=blue,pdftitle={},pdfauthor={},pdfdisplaydoctitle=true,pdfduplex=DuplexFlipLongEdge]{hyperref}

\definecolor{max}{HTML}{03A678}
\definecolor{rafal}{HTML}{161226}

\definecolor{darkgreen}{rgb}{0,0.60,.2}
\definecolor{darkblue}{rgb}{0.1,0.3,1}

\usepackage{physics}            
\usepackage[normalem]{ulem}     
\usepackage[T1]{fontenc}
\usepackage[utf8]{inputenc}
\usepackage{amsthm}
\usepackage{amssymb}
\usepackage{orcidlink}

\usepackage{endnotes}
\let\footnote=\endnote
\date{\today}
\begin{document}

\title{Fading ergodicity and quantum dynamics in random matrix ensembles}

\author{Rafał Świętek\orcidlink{0009-0004-5353-9998}}
\affiliation{Institut f\"ur Theoretische Physik, Georg-August-Universität G\"ottingen, D-37077 G\"ottingen, Germany\looseness=-1}
\affiliation{Department of Theoretical Physics, J. Stefan Institute, SI-1000 Ljubljana, Slovenia}
\affiliation{Department of Physics, Faculty of Mathematics and Physics, University of Ljubljana, SI-1000 Ljubljana, Slovenia\looseness=-1}
\author{Maksymilian Kliczkowski\orcidlink{0000-0002-7987-9913}}
\affiliation{Institute of Theoretical Physics, Faculty of Fundamental Problems of Technology, Wrocław University of Science and Technology, 50-370 Wrocław, Poland\looseness=-3}
\affiliation{Department of Theoretical Physics, J. Stefan Institute, SI-1000 Ljubljana, Slovenia}
\affiliation{Department of Physics, Faculty of Mathematics and Physics, University of Ljubljana, SI-1000 Ljubljana, Slovenia\looseness=-1}
\author{Miroslav Hopjan\orcidlink{0000-0002-0905-1571}}
\affiliation{Institute of Theoretical Physics, Faculty of Fundamental Problems of Technology, Wrocław University of Science and Technology, 50-370 Wrocław, Poland\looseness=-3}
\affiliation{Department of Theoretical Physics, J. Stefan Institute, SI-1000 Ljubljana, Slovenia}
\author{Lev Vidmar\orcidlink{0000-0002-6641-6653}}
\affiliation{Department of Theoretical Physics, J. Stefan Institute, SI-1000 Ljubljana, Slovenia}
\affiliation{Department of Physics, Faculty of Mathematics and Physics, University of Ljubljana, SI-1000 Ljubljana, Slovenia\looseness=-1}
\begin{abstract}
Recent work has proposed fading ergodicity as a mechanism for many-body ergodicity breaking. Here, we show that two paradigmatic random matrix ensembles -- the Rosenzweig–Porter model and the ultrametric model -- fall within the same universality class of ergodicity breaking when embedded in a many-body Hilbert space of spins-1/2. By calibrating the parameters of both models via their Thouless times, we demonstrate that the matrix elements of local observables display similar statistical properties, allowing us to identify the fractal phase of the Rosenzweig–Porter model with the fading-ergodicity regime. This correspondence is further supported through the analyses of quantum-quench dynamics of local observables, their temporal fluctuations and power spectra, and survival probabilities. Our findings reveal that local observables thermalize within the fading-ergodicity regime on timescales shorter than the Heisenberg time, thus providing a unified framework for understanding ergodicity breaking across these distinct models.
\end{abstract}
\maketitle

\section{Introduction\label{sec:intro}}
Determining the boundaries of ergodicity in isolated quantum systems, as well as assessing the possible universality of these boundaries, has emerged as a central challenge in the search for robust non-ergodic phases of matter.
Despite substantial progress in our understanding of quantum thermalization~\cite{dalessio_kafri_16,deutsch_18}, a comprehensive theoretical framework for characterizing systems that depart from ergodic behavior remains elusive~\cite{sierant_manybodylocalizationage_2025}.

Recent work has proposed a mechanism of ergodicity breaking known as fading ergodicity~\cite{kliczkowski_vidmar2024}, inspired by the quantum sun model~\cite{suntajs_vidmar_22} -- a toy model of many-body ergodicity breaking~\cite{deroeck_huveneers_17}.
Fading ergodicity focuses on the behavior of fluctuations in observable matrix elements and provides a framework for formulating a scaling theory of ergodicity breaking with the critical exponent $\nu=1$~\cite{swietek_vidmar_scaling24}.
In this way, it offers a natural conceptual bridge between the standard eigenstate thermalization hypothesis (ETH), introduced by Srednicki~\cite{Deutsch91,srednicki_94,srednicki_99, rigol_dunjko_08, dalessio_kafri_16}, and the full absence of ETH that is characteristic of genuinely nonergodic phases.

An alternative viewpoint to the analysis of physical models with few-body interactions (associated with sparse Hamiltonian matrices) is offered by structured random matrix models.
Such models employ dense Hamiltonian ensembles and frequently permit analytically tractable predictions for the onset of localization in a specified basis.
Among the structured random matrix constructions particularly relevant to this work are the Rosenzweig–Porter model~\cite{rosenzweig_porter1960, kravtsov_khaymovich2015, facoetti_nonergodiceigenvectorslocal_2016, Truong_2016, monthus_multifractalityeigenstatesdelocalized_2017, vonsoosten_nonergodicdelocalizationrosenzweig_2019, bogomolny_sieber_18, buijsman_longrangespectralstatistics_2024, DeTomasi19, skvortsov_kravtsov2022, barney_galitski_23, lin_2023, venturelli_tarzia_23, altland_shapiro97, Pino_2019, backer_khaymovich2019, khaymovich_kravtsov2021, zhang_dietz23, jahnke_kim_25}, the ultrametric model~\cite{fyodorov2009, Rushkin_2011, gustkin_osipov_2011, bogomolny_giraud_2011, Mendez_Bermudez_2012, bogomolny_sieber_18b, soosten_warzel_2017, soosten_warzel_2018, soosten_warzel_2019, suntajs_hopjan_24}, and the power-law random banded model~\cite{mirlin_thomas_96, Evers2000, mirlin_evers_00}.

\begin{figure}[b]
    \centering
    \includegraphics[width=\columnwidth]{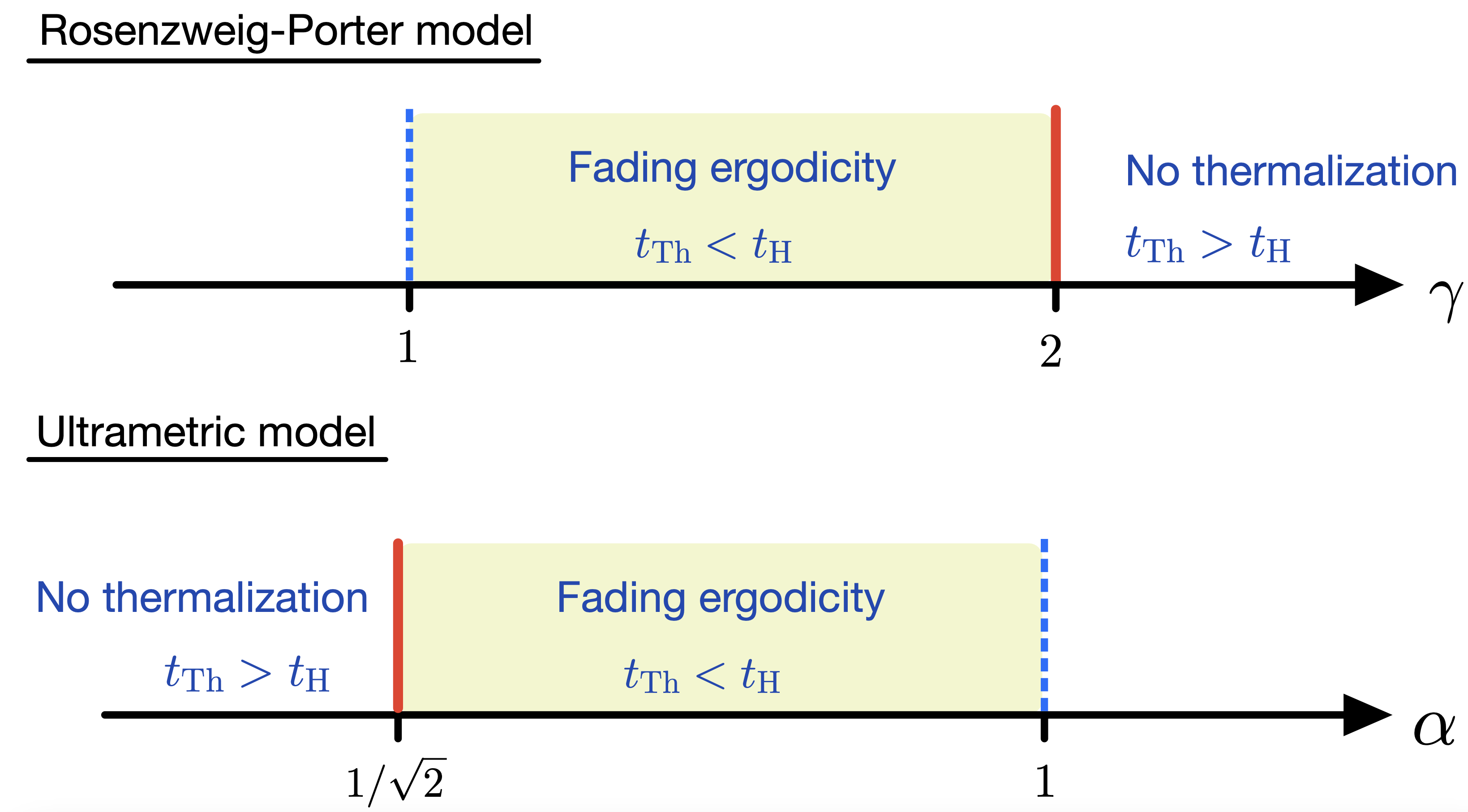}
    \caption{ Sketch of the parameter regimes in both models in which fading ergodicity is observed.
    In the Rosenzweig-Porter model where the tuning parameter is $\gamma$, see Eq.~\eqref{eq:RP model}, fading ergodicity is observed at $1<\gamma<2$, while in the ultrametric model where the tuning parameter is $\alpha$, see Eq.~\eqref{eq:um_def}, fading ergodicity is observed at $1/\sqrt{2}<\alpha<1$.
    In both cases, the observable relaxation time, denoted as the Thouless time $t_{\rm Th}$, is shorter than the Heisenberg time $t_{\rm H}$.
    }
    \label{fig:fig_sketch}
\end{figure}

The central aim of this work is to build a conceptual bridge between many-body ergodicity breaking in physical systems and localization phenomena in random matrix models.
We identify a set of universal features shared by these two classes of systems in the vicinity of the ergodicity-breaking transition, focusing in particular on the fractal phase of the Rosenzweig–Porter model and the ergodic phase of the ultrametric model, see Fig.~\ref{fig:fig_sketch}.
Our analysis shows that the essential characteristics of the precursor regime to ergodicity breaking are naturally captured by the fading ergodicity framework.
This provides a unified viewpoint on ergodicity-breaking mechanisms in both microscopic many-body systems and random matrix ensembles.

To pursue this goal, we first follow a two-step approach. 
(i) We embed the random matrix models into many-body Hilbert spaces composed of $L$ coupled spins-1/2, which allows us to define and examine genuinely local observables, such as $\hat S_j^z$ at site $j$.
(ii) We calibrate the parameters of each random matrix model by aligning their Thouless times $t_{\rm Th}$, such that we find exponentially large (in $L$) relaxation times in this parameter regime~\cite{deroeck_huveneers_17, suntajs_vidmar_22, suntajs_hopjan_24, swietek_vidmar_scaling24}.
With this mapping in place, we study the statistical properties of observable matrix elements and demonstrate that, near the ergodicity-breaking transition, their behavior is well described by the fading ergodicity framework.

We then demonstrate that, within the fading ergodicity regime, local observables display thermalization: following a quantum quench, their expectation values relax toward the predictions of the appropriate statistical ensemble on timescales shorter than the Heisenberg time $t_H$ ($t_H = 2\pi\hbar/\Delta$, where $\Delta$ is the mean level spacing).
To supplement the quench analysis, we examine temporal fluctuations of local observables and their power spectra, as well as the corresponding survival probabilities.
We contrast the behavior of these observables across different random matrix models.

The remainder of this work is structured as follows. In Sec.~\ref{sec:models}, we first introduce the models of interest and review their spectral properties. This allows us to establish connection between their model parameters via the Thouless energy $\Gamma$, which is proportional to the inverse of the Thouless time $t_{\rm Th}$.
In Sec.~\ref{sec:anomalous_dynamics}, we show how the Thouless energy can be extracted from the spectral function of observables that exhibit fading ergodicity.
In Sec.~\ref{sec:quench}, we then study quantum dynamics of the same observable after a quench from initial product states. 
It exhibits a short-time exponential decay with the rates given by the corresponding Thouless energies, and in Sec.~\ref{sec:thermalization} we show that their long-time averages approach the microcanonical ensemble predictions in the thermodynamic limit.
The variance of temporal fluctuations is studied in Sec.~\ref{sec:fluctuations}, and their power spectra and the survival probability are studied in Sec.~\ref{sec:noise}.
We conclude in Sec.~\ref{sec:conclusion}.


\section{Set-up and Models\label{sec:models}}

The physical system that we have in mind consists of $L$ qubits or spins-1/2.
The corresponding many-body Hilbert space ${\cal H}^{(L)}$ is then a tensor product of $L$ local Hilbert spaces ${\cal H}_i$ with dimensions $d = {\rm dim}({\cal H}_i)=2$, i.e., ${\cal H}^{(L)} =\otimes_{i=1}^L {\cal H}_i$, and the dimension of the many-body Hilbert space is ${\cal D} = {\rm dim}({\cal H}^{(L)})=2^L$.
The basis states are referred to as the computational basis, 
$\ket{\sigma_1\cdots\sigma_L} \equiv \ket{\sigma}_1 \otimes \ket{\sigma}_2 \otimes ... \otimes \ket{\sigma}_L$, where $\ket{\sigma}_i = \{ \ket{\uparrow}_i, \ket{\downarrow}_i \}$ are the basis states of a single qubit.

All the Hamiltonians $\hat H$ and the operators $\hat O$ studied in this work are defined in this many-body Hilbert space.
For the structured random matrix models, $\hat H$ is represented by dense matrices, which, in this particular many-body Hilbert space, indicates presence of multi-body (unphysical) interactions.
Yet, recent work argued that properties of ergodicity breaking transitions in the quantum sun model~\cite{deroeck_huveneers_17, suntajs_vidmar_22}, which is described by few-body (physical) interactions and hence being represented by sparse Hamiltonian matrices, share striking similarities with those in the ultrametric model of random matrix ensembles~\cite{suntajs_hopjan_24}.
Therefore, the main goal of this work is to establish a common mechanism of ergodicity breaking transition in structured random matrix models, when defined in the many-body Hilbert space of coupled spins-1/2, and in certain physical models such as the quantum sun model.


\subsection{Rosenzweig-Porter (RP) model\label{sec:models:RP_model}}

The RP model is a rather simple random matrix model in which the standard deviation of the off-diagonal matrix elements is suppressed when compared to the diagonal matrix elements.
We define the RP model Hamiltonian as~\cite{kravtsov_khaymovich2015}
\begin{equation} \label{eq:RP model}
    \hat{H}=\hat{H}_0+\frac{1}{\mathcal{D} ^{\gamma/2}}\hat{M}\;,
\end{equation}
where $\hat{H}_0$ is a diagonal operator and $\hat M$ is an operator that includes off-diagonal terms.
The diagonal matrix elements of $\hat H_0$ are drawn from a normal distribution with zero mean and unit variance, and $\hat{M}$ is defined as $\hat{M}=(\hat A + \hat A^\dagger)/\sqrt{2}$, 
with all matrix elements of $\hat{A}$ drawn from a normal distribution with zero mean and unit variance.
Hence, $\hat M$ also contains nonzero diagonal matrix elements, which is the definition also used in several previous works~\cite{Kunz1998,Truong_2016, venturelli_tarzia_23, barney_galitski_23, cadez_dietz_2024, buijsman_longrangespectralstatistics_2024, Kutlin2025}.

Properties of the model are controlled by the parameter $\gamma$.
At $\gamma < 1$, the spectral properties match those of the Gaussian orthogonal ensemble (GOE)~\cite{kravtsov_khaymovich2015, facoetti_nonergodiceigenvectorslocal_2016, monthus_multifractalityeigenstatesdelocalized_2017, vonsoosten_nonergodicdelocalizationrosenzweig_2019} and the system is ergodic.
In the opposite limit, $\gamma>2$, the energy spectrum follows the Poisson level statistics and the eigenstates are localized in a given basis~\cite{vonsoosten_nonergodicdelocalizationrosenzweig_2019}.

In this work, we focus on the intermediate regime $1<\gamma<2$, in which short-range spectral statistics flow towards the GOE limit~\cite{Pino_2019}, while long-range statistics do not follow the prediction of a finite size GOE~\cite{hopjan_vidmar_23b, buijsman_longrangespectralstatistics_2024}, see Appendix~\ref{sec:gap_ratio}.
This regime is also referred to as the ``fractal phase'' since Hamiltonian eigenstates exhibit a fractal structure with the fractal dimension of $d_q^{\rm(eig)}=2-\gamma$ for $q>1/2$~\cite{kravtsov_khaymovich2015, Truong_2016, bogomolny_sieber_18, DeTomasi19, vonsoosten_nonergodicdelocalizationrosenzweig_2019, deluca2014, detomasi_multifractalitymeetsentanglement_2020}.
We define the ergodicity breaking transition point in the RP model at $\gamma=\gamma_c = 2$, since at this value of $\gamma$ the Thouless time $t_{\rm Th}$ crosses the Heisenberg time $t_{\rm H}$, see Sec.~\ref{sec:models:calibrating} for further details.
Correspondingly, the point $\gamma=2$ is associated with the complete disappearance of the GOE statistics with a critical exponent $\nu=1$~\cite{Pino_2019, khaymovich_kravtsov2021}.


\subsection{Ultrametric (UM) model\label{sec:models:UM}}

The UM model is a hierarchical random matrix model that describes, when expressed in the basis of $L$ qubits, the interplay of $N$ central impurities interacting with the remaining $L'=L-N$ qubits via a spatially decaying coupling~\cite{suntajs_hopjan_24}.
We define the UM model Hamiltonian as
\begin{equation} \label{eq:um_def}
    \hat{H} = \hat{H}_0 + J\sum _{k=1}^{L'} \alpha ^k \hat{H}_k\;,
\end{equation}
where $\hat{H}_k$ is a block-diagonal random matrix.
In the matrix representation, $H_k$ is a direct sum of $2^{L'-k}$ independently sampled random matrices of size $2^{N+k} \times 2^{N+k}$, i.e.,
\begin{equation}
    H_k = \bigoplus_{i=1}^{2^{L'-k}} H_k^{(i)}\;,\;\;\;
    H_k^{(i)} = \frac{1}{\sqrt{2^{N+k} + 1}}R^{(i)}\;, 
\end{equation}
with $R^{(i)}$ sampled from the GOE with dimension $2^{N+k}\times2^{N+k}$, and the prefactor is due to normalization.
The parameter $\alpha \leq 1$ in Eq.~\eqref{eq:um_def} determines the exponential decay of the coupling with distance $k$. Unless otherwise specified, we fix $J=1$ and $N=1$, with similar results observed for other values of $N$. 

The UM model, defined in the many-body Hilbert space, undergoes an ergodicity breaking phase transition at $\alpha_c=1/\sqrt{2}$~\cite{suntajs_hopjan_24}.
At the critical point $\alpha_c$, the model exhibits multifractal eigenstates~\cite{suntajs_hopjan_24} with fractal dimensions $d_q^{\rm(eig)}$ varying with the index $q > 0$.
In the ergodic phase, $\alpha > \alpha_c$, the short-range spectral statistics flow towards the GOE limit~\cite{suntajs_hopjan_24}, while long-range statistics do not follow the prediction of a finite size GOE~\cite{hopjan_vidmar_23b}. Here, the eigenstates are at most weakly multifractal, with fully ergodic ones ($d_q^{\rm(eig)}\approx1$) deep in the ergodic phase at $\alpha \approx1$~\cite{suntajs_hopjan_24}.

While the focus of this work is on the UM model, we note that similar behavior is expected to emerge in the power-law random banded (PLRB) model~\cite{mirlin_thomas_96}, if expressed in the same many-body Hilbert space.
The PLRB model is most commonly thought of as a single-particle model with long-range hopping, in which the standard deviation of a typical off-diagonal matrix element $h_{i,j}$ decays with distance $r$ as ${\rm std}(h_{i,j}) \propto r^{-a}$, where $a$ is the control parameter.
When the UM and PLRB models are expressed in the same Hilbert space, the relationship between the parameters $a$ of the PLRB model and $\alpha$ of the UM model is~\cite{suntajs_hopjan_24}
\begin{equation} \label{eq:um_plrbm}
a=1/2-\log_2 \alpha\;.
\end{equation}
While the implementation of the PLRB model in the many-body Hilbert space has been recently discussed in~\cite{bujisman_khaymovich_25}, the actual study of the observable matrix elements and quantum dynamics is left for future work.


\subsection{Calibrating the model parameters\label{sec:models:calibrating}}

The RP and UM models exhibit two important energy scales.
The first is the mean level spacing $\Delta$ that is, in our set-up, exponentially small in $L$.
In the actual numerical calculations we define $\Delta$ as $\Delta=\expval{E_{n+1}-E_n}_n$, where $E_n$ are the Hamiltonian eigenenergies and the average $\expval{...}_n$ is taken over a narrow window of $500$ mid-spectrum eigenstates.
Another energy scale is the Thouless energy $\Gamma$, which also gives rise to the definition of the corresponding Thouless time $t_{\rm Th} = 2\pi\hbar/\Gamma$ (we set $\hbar\equiv 1$ further on).
A convenient approach to extract the Thouless time $t_{\rm Th}$ is via the spectral form factor~\cite{Bertini2018, Kos2018, Friedman2019, Vasilyev2020, Winer2020, Liao2020, prakash_pixley_21, suntajs_bonca_20a, suntajs_prosen_21, sierant_delande_20, Colmenarez2022, suntajs_vidmar_22, Joshi2022, Winer2022, barney_galitski_23, buijsman_longrangespectralstatistics_2024, Ceven2026}, in which $t_{\rm Th}$ is defined as the onset time of the universal  GOE ramp~\cite{berry_85, cotler_chaoscomplexityrandom_2017, Gharibyan2018, Chan2018}.
In this work, instead, we extract the Thouless energy $\Gamma$ via the width of the spectral function of a given observable, to be discussed in Sec.~\ref{sec:Eth_spectral}.
If one chooses the observable for which its dynamical response corresponds to the slowest relaxation process in the system, we expect that $1/\Gamma$ will correspond to $t_{\rm Th}$ extracted from the spectral form factor.
This expectation follows from the intuition that the Thouless time represents the longest physically relevant relaxation time in the system.

Another convenient property of both models is that accurate information about the Thouless energy can be obtained via the rate of the Fermi golden rule.
In the RP model, it scales as~\cite{bogomolny_sieber_18, DeTomasi19, skvortsov_kravtsov2022, barney_galitski_23, lin_2023, venturelli_tarzia_23, buijsman_longrangespectralstatistics_2024}
\begin{equation}\label{eq:RP:thouless}
    \Gamma_{\rm RP} \propto \pi\mathcal{D}^{1-\gamma}\propto (2^L)^{1-\gamma}\,,
\end{equation}
while in the UM model, it scales as~\cite{luitz_huveneers_17, suntajs_vidmar_22, suntajs_hopjan_24}
\begin{equation}\label{eq:UM:thouless}
    \Gamma_{\rm UM} \propto \alpha^{2L'}\;.
\end{equation}
Note that in Eq.~\eqref{eq:UM:thouless}, one may replace $L'$ with $L$, since $L=L'+N$ and $N$ is a constant, when approaching the thermodynamic limit.

The key property of the Thouless energies in both models, cf.~Eqs.~\eqref{eq:RP:thouless} and~\eqref{eq:UM:thouless}, is that they are exponentially small in $L$ at $\gamma>1$ and $\alpha<1$, respectively.
Hence, at some values of $\gamma$ and $\alpha$, the Thouless energy $\Gamma$ scales identically with $L$ as the mean level spacing $\Delta$, which corresponds to the Heisenberg energy.
This occurs at $\gamma=\gamma_c = 2$ in the RP model and at $\alpha=\alpha_c=1/\sqrt{2}$ in the UM model.
The criterion $\Gamma \sim \Delta$ determines the ergodicity breaking transition point, and therefore, we refer to $\gamma_c$ and $\alpha_c$ as the critical points of the ergodicity breaking phase transition.

In what follows, we calibrate the parameters of both models such that they correspond to the same Thouless time, or, equivalently, to the same Thouless energy,
\begin{equation}\label{eq:RP model_UM_th_match}
    \Gamma_{\rm UM} = \Gamma_{\rm RP}\;\;\;\rightarrow\;\;\; c_{\rm UM} \alpha^{2L}= c_{\rm RP} 2^{L(1-\gamma)}\;,
\end{equation}
where the constants $c_{\rm RP}$ and $c_{\rm UM}$, not explicitly written out in Eqs.~\eqref{eq:RP:thouless} and~\eqref{eq:UM:thouless}, respectively, may contain a subleading (polynomial) dependence on $L$.
Equation~\eqref{eq:RP model_UM_th_match} gives rise to the relationship between the parameters $\alpha$ and $\gamma$,
\begin{equation}\label{eq:RP model_UM_relation:finiteL}
    \gamma= 1+\frac{\ln{\alpha}}{\ln{\alpha_c}}+\frac{\ln(c_{\rm RP}/c_{\rm UM})}{L\ln{2}}\;,
\end{equation}
in which we replaced $1/\sqrt{2}$ by $\alpha_c$.
The last term on the r.h.s.~of Eq.~\eqref{eq:RP model_UM_relation:finiteL} is a finite-size correction, whose quantitative impact will be tested in Sec.~\ref{sec:Eth_spectral}.
In the thermodynamic limit $L\rightarrow \infty$, Eq.~\eqref{eq:RP model_UM_relation:finiteL} simplifies to
\begin{equation}\label{eq:RP model_UM_relation}
    \gamma= 1+\frac{\ln{\alpha}}{\ln{\alpha_c}}\;,
\end{equation}
which is our reference point when calibrating the parameters of both models.
Equation~\eqref{eq:RP model_UM_relation} is relevant in the regime $\gamma>1$, $\alpha<1$, which is the regime of our interest.
We note that in the PLRB model, one can connect the parameter $a$ to $\gamma$ as $a=\gamma/2$.

In the remainder of the paper, we numerically study properties of the observable matrix elements and quantum quench dynamics in both models.
A single Hamiltonian realization is denoted as $\hat H^{(\mu)}$ and the average over Hamiltonian realization is denoted as ${\rm Avr}_\mu\{\cdots\}$.
Our results correspond to averages over $N_{\rm r}$ Hamiltonian realizations.
In the RP model, we set $N_{\rm r}=7500$ for $L\leq12$ and $N_{\rm r}\geq3000, 1200, 600, 50$ for $L = 13, 14, 15, 16$, respectively.
In the UM model, we set $N_{\rm r}\gtrsim 1500$ for $L\leq 11$, $N_{\rm r}\gtrsim 2500$ for $L = 12$, $N_{\rm r} \gtrsim 1500$ for $L=13, 14$, $N_{\rm r}\gtrsim 400$ for $L=15$, and $N_{\rm r}\lesssim 100$ for $L=16$.

\section{Fluctuations of matrix elements\label{sec:anomalous_dynamics}}

\subsection{Conventional ETH and Fading ergodicity} \label{sec:eth_fading}

We start by reviewing the theoretical frameworks relevant for the description of the behavior of observable matrix elements.
We study the expectation values of an observable $\hat{O}$ in the eigenbasis of a Hamiltonian, $\smash{\{\ket{n}: \hat{H}\ket{n} = E_n\ket{n}\}}$.
The conventional ETH states that the matrix elements $\langle n | \hat{O} | m \rangle$ satisfy the Srednicki ansatz~\cite{Deutsch91, srednicki_94, rigol_dunjko_08, dalessio_kafri_16},
\begin{equation}\label{eq:ETH}
    \langle n | \hat{O} | m \rangle = O (\bar E) \delta_{nm} + \rho (\bar E) ^ {-1/2} f(\bar E, \omega _{nm}) R_{nm}\;,
\end{equation}
where $\bar E = (E_n + E_m)/2$ represents the average energy, $\omega_{nm} = E_n - E_m$ is the energy difference, and $R_{nm}$ is a random variable with zero mean and unit variance.
The functions $O (\bar E)$ and $f(\bar E, \omega_{nm})$ are smooth functions of their arguments.
The suppression of matrix elements fluctuations is governed by the many-body density of states $\rho(\bar E)$, which scales as $\rho(\bar E) \propto 2^L$ for states in the middle of the spectrum.
The validity of Eq.~\eqref{eq:ETH} was numerically tested in several ergodic quantum many-body systems~\cite{rigol_09a, rigol_santos_10, rigol09, santos_rigol_10a, santos_rigol_10b, steinigeweg_herbrych_13, khatami_pupillo_13, sorg14, beugeling_moessner_14, steinigeweg_khodja_14, mondaini_fratus_16, dymarsky_lashkari_18, yoshizawa_iyoda_18, jansen_stolpp_19, richter_dymarsky_20, leblond_rigol_20, mierzejewski_vidmar_20, santos_perezbernal_20, brenes_goold_20, brenes_leblond_20, schoenle_jansen_21, Dymarsky2022, wang_lamann_22, capizzi_2024, foini_kurchan_19, pappalardi_foini_22, Pappalardi2025}.

In contrast to the conventional ETH, fading ergodicity accounts for modifications of Eq.~\eqref{eq:ETH} when approaching the ergodicity breaking transition.
As argued in Ref.~\cite{kliczkowski_vidmar2024}, the divergence of $t_{\rm Th}$ gives rise to the accumulation of the spectral weight at $\omega \approx 0$, and hence the suppression of the matrix elements fluctuations in Eq.~\eqref{eq:ETH} is no longer given by $\rho (\bar E)^{-1/2}$.
In the fading ergodicity regime, we express the average off-diagonal matrix elements at low-frequency, $\omega\ll\Gamma$, as
\begin{equation}\label{eq:fading}
    \text{Avr}\left[\abs{\langle n | \hat{O} | m \rangle}^2\right]_{\bar{E},\omega\ll\Gamma}\propto\frac{1}{\rho\Gamma}\propto\rho^{-2/\eta}\,,
\end{equation}
where $\text{Avr}\left[...\right]$ denotes the average over the states $| n\rangle \neq |m\rangle$ within a microcanonical window centered at $\bar E$. 
In Eq.~\eqref{eq:fading} we introduced the exponent $\eta$ that varies continuously between the value $2$ in the conventional ETH, see Eq.~\eqref{eq:ETH}, and the ergodicity breaking critical point at which it diverges.
Physically, the growing exponent $\eta$ indicates a gradual loss of ergodicity as the weight of the matrix elements concentrates at $\omega\approx 0$, and hence it provides a bridge between the conventional ETH and the ergodicity breaking transition.

\subsection{Thouless energy from the spectral function} \label{sec:Eth_spectral}

The discussion in Sec.~\ref{sec:eth_fading} highlights the need to obtain accurate information about the scaling of Thouless energy $\Gamma$.
We next describe how to extract $\Gamma$ from the spectral function $\tilde{\mathcal{O}}(\omega)$ of observables of interest.
The spectral function corresponds to the Fourier transform of the connected eigenstate autocorrelation function $\tilde{\mathcal{O}}(t)$, defined as
\begin{equation} \label{def_Ot_allstates}
    \tilde{\mathcal{O}}(t) = \frac{1}{\cal D} \sum_{n=1}^{\cal D} \left( \langle n|\hat O(t)\hat O|n\rangle - \langle n|\hat O(t)|n\rangle \langle n|\hat O|n\rangle \right)\;,
\end{equation}
such that the spectral function is
\begin{equation}
    \tilde{\mathcal{O}}(\omega) = \int_{-\infty}^\infty dt e^{i\omega t} \, \tilde{\mathcal{O}}(t) = \frac{1}{\mathcal{D}}\sum_{n\neq m}\abs{O_{nm}}^2\delta(\omega-\omega_{nm})\;,
\end{equation}
where $O_{nm}=\langle n|\hat{O}|m\rangle$.
In the actual numerical calculations, we use the coarse-grained version of the spectral function for a fixed Hamiltonian realization $\hat H^{(\mu)}$ as
\begin{equation}\label{eq:f_function:coarse:mu}
    \mathcal{O}^{(\mu)}(\omega) = \hspace*{-0.5cm} \sum_{\substack{n\neq m:\\ |\omega_{nm}-\omega|<\delta\omega/2}} \hspace*{-0.5cm} \frac{\abs{O_{nm}}^2}{\mathcal{M'_\omega}}\,.
\end{equation}
The sum in Eq.~\eqref{eq:f_function:coarse:mu} is carried over a bin of width $\delta\omega$ and $\mathcal{M}'_\omega=\sum_{\substack{n\neq m:\ |\omega_{nm}-\omega|<\delta\omega/2}}$ is the number of elements in the bin.
So far, the Hamiltonian eigenstates and the observables matrix elements referred to a single Hamiltonian realization $\hat H^{(\mu)}$, i.e., $|n\rangle \equiv |n^{(\mu)}\rangle$ and $O_{nm} \equiv O_{nm}^{(\mu)}$.
We subsequently average the spectral function over Hamiltonian realizations to obtain
\begin{equation}\label{eq:f_function:coarse}
    \mathcal{O}(\omega)={\rm Avr}_\mu\qty{\mathcal{O}^{(\mu)}(\omega)}\;.
\end{equation}
The bins of the coarse-graining procedure in Eq.~\eqref{eq:f_function:coarse:mu} are logarithmically spaced between $0.1/\mathcal{D}$ and the bandwidth $\Delta E$, see Appendix~\ref{sec:app:variance} for the definition of $\Delta E$.
The bins are identical for all Hamiltonian realizations. 

\begin{figure}[t]
    \centering
    \includegraphics[width=\columnwidth]{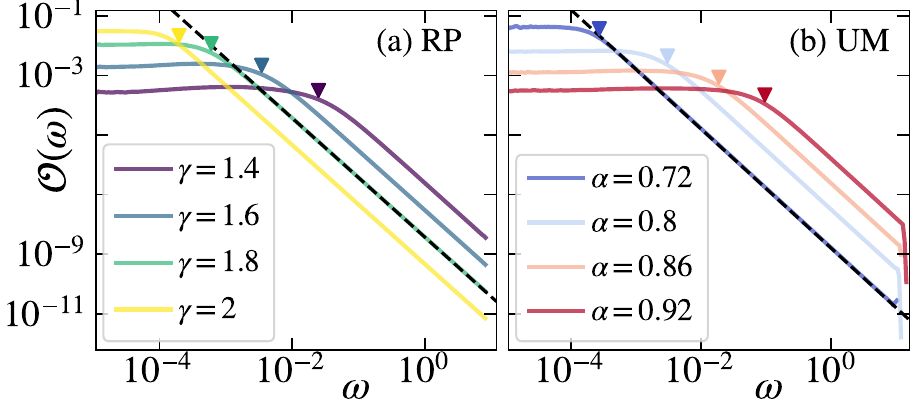}
    \caption{Coarse grained spectral function $\mathcal{O}(\omega)$, see Eq.~\eqref{eq:f_function:coarse}, of the observable ${\hat{O} = \hat{S}^z_L}$ and $L=15$ for (a) the RP model at different $\gamma$ and (b) the UM model at different $\alpha$.
    The symbols denote the position of the Thouless energy extracted using Eq.~\eqref{eq:integrated_spec}. The dashed lines represent a characteristic $1/\omega^2$ decay.}
    \label{fig:fig_O}
\end{figure}

Figure~\ref{fig:fig_O} shows examples of spectral functions $\mathcal{O}(\omega)$ for the observable $\smash{\hat{O} = \hat{S}^z_L}$.
In the UM model, the qubit at $i=L$ corresponds to the qubit that is most weakly coupled to the central impurity, while in the RP model, the choice of the qubit is not essential.
In Figs.~\ref{fig:fig_O}(a) and~\ref{fig:fig_O}(b), we observe a very similar behavior of the spectral functions in both models, with a plateau at small $\omega$, followed by a Lorentzian decay ($\propto 1/\omega^{2}$) at large $\omega$.

The Lorentzian shape of the spectral functions in Fig.~\ref{fig:fig_O} is similar to the spectral function of the same observable in the quantum sun model~\cite{kliczkowski_vidmar2024}.
This is an indication about the common mechanism of the ergodicity breakdown in these models.
However, the most important ingredient in our analysis is not the shape of the spectral function, but rather the choice of the observable $\hat S_L^z$.
Choosing the qubit at site $i=L$ in the UM model guarantees that its relaxation time will be longer than the relaxation times of other qubits, and hence one expects its relaxation time to correspond to the Thouless time $t_{\rm Th}$.
In the energy domain, which is relevant for the spectral functions, we then associate the width of the plateau of the spectral function with the Thouless energy $\Gamma$.
The latter shrinks to the Heisenberg energy $\Delta$ at the ergodicity breaking transition point, and the regime in which $\Gamma$ approaches $\Delta$, but is still larger than $\Delta$, is referred to as the fading ergodicity regime.
\begin{figure}[t]
    \centering
    \includegraphics[width=0.9\columnwidth]{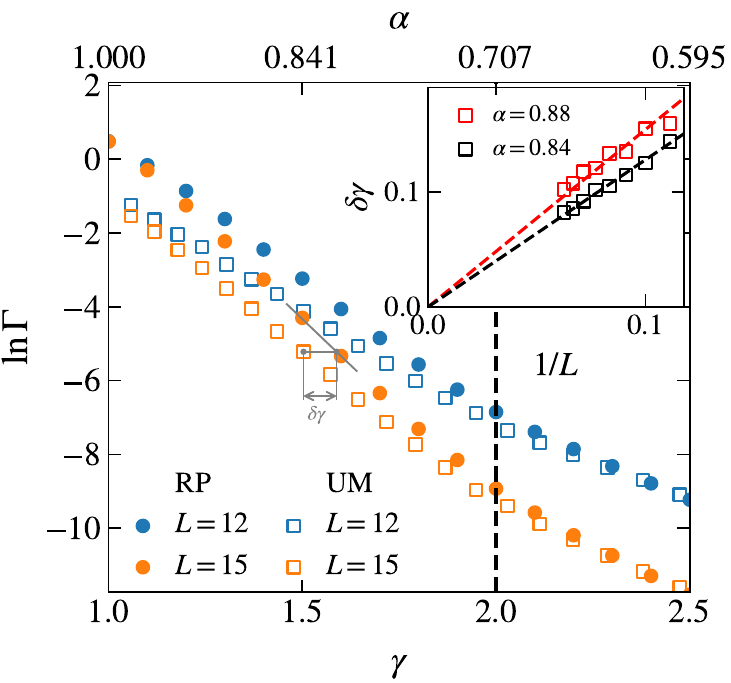}
    \caption{
        The logarithm of the Thouless energy $\Gamma$
        for the UM [RP] model as function of control parameter
        $\alpha$ [$\gamma$] for two system sizes $L$.
        We determine the distance $\delta\gamma$, see the sketch in the main panel, by calculating the deviation of the numerical result for the UM model from a linear fit to the RP model results within the same range of $\Gamma$.
        Inset: Finite size scaling of $\delta\gamma$, extracted from the main panel, versus $1/L$, for two values of $\alpha$.
        Dashed lines are fits of the linear function to the results.
    }\label{figM1}
\end{figure}

For a quantitative extraction of the Thouless energy $\Gamma$ from $\mathcal{O}(\omega)$, we follow the procedure of Refs.~\cite{kliczkowski_vidmar2024, swietek_lydzba25}.
We first introduce the integrated spectral function,
\begin{equation}\label{eq:integrated_spec}
    I_{\mathcal{O}}(\omega)=\int_0^\omega d\omega\mathcal{O}(\omega)\;,
\end{equation}
which we normalize to the interval $[0,1]$.
We find the Thouless energy $\Gamma$ as a point at which $I_{\mathcal{O}}(\omega=\Gamma)=1/2$.
While this procedure works best for the Lorentzian form of the spectral functions, it also provides reliable results when small deviations from the Lorentzian functions are observed~\cite{swietek_lydzba25}.

We next test whether the Thouless energies $\Gamma$, extracted from the widths of the spectral functions as described above, match the predicted relation in Eq.~\eqref{eq:RP model_UM_relation}.
We show the extracted values of $\Gamma$ in Fig.~\ref{figM1} as a function of parameter $\gamma$ [$\alpha$] for the RP [UM] model. The values are obtained at two system sizes $L=\{12,15\}$. The vertical dashed line denotes the critical points $\gamma_c$ and $\alpha_c$.
The agreement between the two models is reasonably good, but, at finite $L$, it is not perfect.
We nevertheless find that qualitative agreement between the Thouless energies is restored in the thermodynamic limit $L\to\infty$.
The restoration can be demonstrated by taking into account the correction $\delta\gamma$ arising from Eq.~\eqref{eq:RP model_UM_relation:finiteL},
\begin{equation}\label{eq:RP model_UM_relation:correction}
    \delta\gamma=\frac{\ln(c_{\rm RP}/c_{\rm UM})}{L\ln{2}}\;.
\end{equation}
In particular, we estimate the distance $\delta\gamma$ between the results for the UM model and the associated curve obtained for the RP model; see also the sketch in Fig.~\ref{figM1} that shows how we extract $\delta\gamma$.
Remarkably, we find that $\delta\gamma$ eventually disappears in the thermodynamic limit, validating the relation in Eq.~\eqref{eq:RP model_UM_relation}.
In the inset of Fig.~\ref{figM1}, we fit the function $\propto 1/L$ to the results for $\delta\gamma$ at two values of $\alpha$, thereby providing relevance of the prediction in Eq.~\eqref{eq:RP model_UM_relation:correction} to describe the finite-size corrections to Eq.~\eqref{eq:RP model_UM_relation}.

\begin{figure}[t]
    \centering
    \includegraphics[width=\columnwidth]{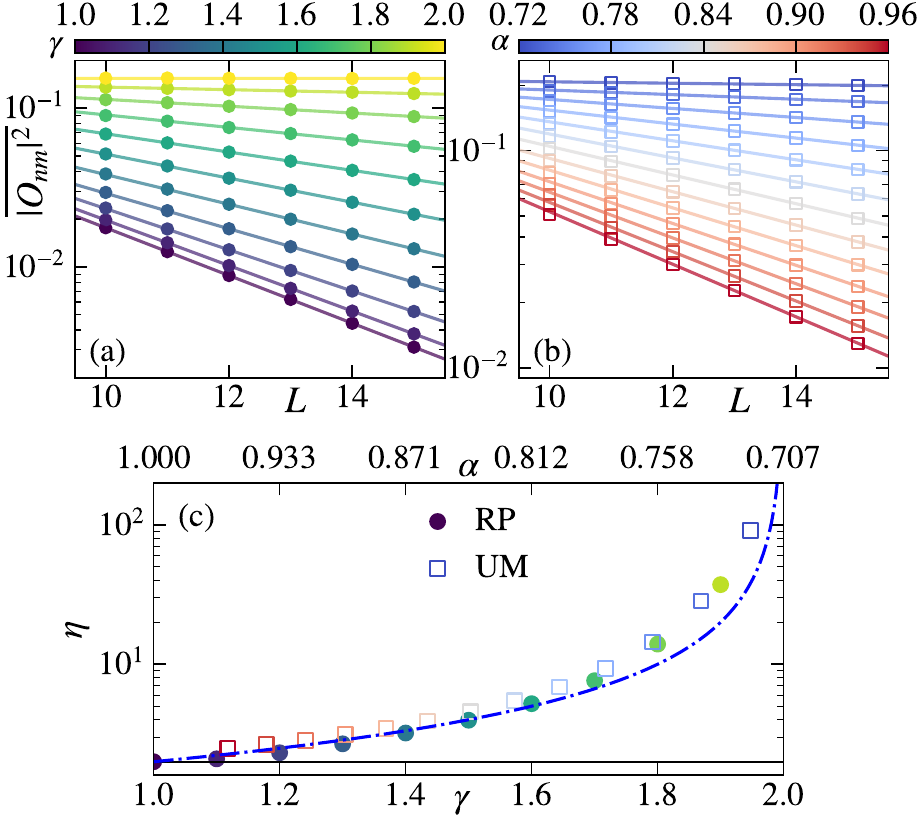}
    \caption{
    Scaling of the off-diagonal matrix elements with the system size $L$ for the (a) RP model and (b) UM model in the frequency range $\smash{\omega\in\qty[\sqrt{\Gamma\Delta}/2,2\sqrt{\Gamma\Delta}]}$. The solid lines show fits of the function $\propto 2^{2L/\eta}$ to the results.
    (c) Fluctuation exponent $\eta$ extracted from the fits in panels (a,c) for the RP model (circles) and UM model (squares) as function of $\gamma$ [$\alpha$] on the lower [upper] x-axis. The parameters $\alpha$ and $\gamma$ are related through Eq.~\eqref{eq:RP model_UM_relation}.
    The dash-dotted line denotes the analytical prediction for $\eta$ from Eq.~\eqref{eq:eta:prediction}.
    }\label{figM_eta}
\end{figure}

\subsection{Extraction of the exponent $\eta$}

After establishing the relationship between the parameters $\gamma$ and $\alpha$, we proceed by demonstrating that the exponent $\eta$ from Eq.~\eqref{eq:fading} behaves qualitatively similar in both models.

We numerically extract $\eta$ by studying the decay of the off-diagonal matrix elements at low $\omega$.
We average the off-diagonal matrix elements around the frequency $\omega=\sqrt{\Gamma\Delta}$, such that the targeted $\omega$ corresponds to the plateau of the spectral function. 
The corresponding average matrix elements $\overline{|O_{nm}|^2}$ are shown vs $L$ in Figs.~\ref{figM_eta}(a) and~\ref{figM_eta}(b) for the RP and UM models, respectively.
Following Eq.~\eqref{eq:fading}, we fit their decay with $L$ using the function $\overline{|O_{nm}|^2} \propto \mathcal{D}^{-2/\eta} = 2^{-2L/\eta}$, yielding good agreement.
We note that $\overline{|O_{nm}|^2}$ corresponds to both the average over a narrow interval around the target $\omega$ for a fixed Hamiltonian realization, as well as the subsequent average over Hamiltonian realizations.

The extracted values of $\eta$ are shown in Fig.~\ref{figM_eta}(c) for the RP [UM] model as a function of $\gamma$ [$\alpha$].
In the fading ergodicity regime, we find remarkable agreement of the estimated $\eta$'s with the analytical expression,
\begin{equation}\label{eq:eta:prediction}
    \eta=2\qty(1-\frac{\ln\alpha}{\ln\alpha_c})^{-1}=\,\frac{2}{2-\gamma}\;,
\end{equation}
where the first equality was established in the study of the quantum sun model~\cite{kliczkowski_vidmar2024}, and the second equality follows from Eq.~\eqref{eq:RP model_UM_relation}.
Equation~\eqref{eq:eta:prediction}, which is free of any fitting parameter, is represented by the dashed-dotted line in Fig.~\ref{figM_eta}(c).
It exhibits a reasonably good agreement with the numerical results for $\eta$.
Similar observations apply when considering the diagonal matrix elements instead of the off-diagonal ones, see Appendix~\ref{sec:app:fading:diagonal} for details. 
Moreover, we note that indicators of spectral statistics, such as the average gap ratio $r$, can also be used to determine the exponent $\eta$.
We demonstrate that in Appendix~\ref{sec:gap_ratio}.

The results in Fig.~\ref{figM_eta} establish the first main result of this work.
They highlight similar behavior of the diverging exponent $\eta$ upon approaching the ergodicity breaking transition, and suggest similarities between the random matrix models studied in this work and the previously studied quantum sun model~\cite{kliczkowski_vidmar2024}.

We finish the section by adding two comments.
First, in the RP model one can relate the parameter $\gamma$ to the fractal dimension $d_2^{\rm (eig)}$ of the Hamiltonian eigenstate $|n\rangle$ in the computational basis.
The latter is calculated from the scaling of the inverse participation ration (IPR) as
\begin{equation} \label{def_ipr_eig}
    {\rm IPR}^{\rm (eig)} = \sum_{\sigma_1,\sigma_2,...,\sigma_L} |\langle \sigma_1\sigma_2...\sigma_L|n\rangle|^4 \propto {\cal D}^{-d_2^{\rm (eig)}} \;.
\end{equation}
Using the known relation $d_2^{\rm (eig)} = 2-\gamma$~\cite{kravtsov_khaymovich2015, vonsoosten_nonergodicdelocalizationrosenzweig_2019, DeTomasi19, khaymovich_kravtsov2021}, it follows from Eq.~\eqref{eq:eta:prediction} that
\begin{equation}
    \frac{2}{\eta} = d_2^{\rm (eig)}\;.
\end{equation}
This equation connects the fluctuations of matrix elements of the observable $\hat S_L^z$ in the RP model with the structure of Hamiltonian eigenstates in the computational basis.
The second comment concerns the choice of the observable under investigation, $\hat O = \hat S_L^z$.
This observable is sensitive to the ergodicity breaking transition since its eigenbasis is the computational basis in which Hamiltonian eigenstates are localized deep in the nonergodic phase~\cite{suntajs_hopjan_24}.
Moreover, choosing the site of the qubit $i=L$ (relevant for the UM model) allows us to associate its relaxation time to the Thouless time, as discussed in Sec.~\ref{sec:Eth_spectral}.
Other observables that are expected to exhibit similar behavior are their products, such as $\hat S_{L-1}^z\hat S_L^z$~\cite{kliczkowski_vidmar2024}.
While the focus of this work are few-body observables, it would be interesting to extend the study of fading ergodicity to entanglement entropies in the RP and UM models~\cite{detomasi_multifractalitymeetsentanglement_2020, bujisman_khaymovich_25}, which is left for future work.


\section{Quantum quench dynamics\label{sec:quench}}

We now focus on the quantum quench dynamics of observables.
We consider the same observable as in Sec.~\ref{sec:anomalous_dynamics}, $\hat O = \hat S_L^z$, since its sensitivity to localization yields the dynamics nontrivial.

The Lorentzian shape of the spectral functions shown in Fig.~\ref{fig:fig_O} suggest exponential decays of the observable autocorrelation functions, with the rate that should correspond to the Thouless energy $\Gamma$.
Matching the values of $\Gamma$ in both models via Eq.~\eqref{eq:RP model_UM_relation}, we expect similar dynamical behavior.

Nevertheless, in this section we do not consider the dynamics of infinite-temperature autocorrelation functions as given by Eq.~\eqref{def_Ot_allstates}, but we study simple, experimentally accessible, initial states $|\psi_0\rangle$ that are product states in computational basis,
\begin{equation}
    |\psi_0\rangle = |\sigma_1\sigma_2...\sigma_L\rangle\;,
\end{equation}
i.e., they are eigenstates of the observable $\smash{\hat{S}^z_L}$.
To minimize finite-size effects, we select, for each individual Hamiltonian realization $\hat{H}^{(\mu)}$, the state $|\psi_0\rangle$ whose energy is closest to the mean energy,
$\mel{\psi_0}{\hat{H}^{(\mu)}}{\psi_0} = \epsilon _ {\psi _0} \approx{E}^{(\mu)}_{\rm av} \equiv \Tr \{\hat{H}^{(\mu)}\} / \mathcal{D}$.
The evolution of the observable $\hat{S}^z_L$ is defined as
\begin{equation}\label{eq:quench}
    Q^{(\mu)}(t) = \mel{\psi_0}{\hat{S}^z_L(t)}{\psi_0}\;,
\end{equation}
and the autocorrelation function is
\begin{equation}\label{eq:autocorrelation}
    C^{(\mu)}(t)=\mel{\psi_0}{\hat{S}^z_L(t)\hat{S}^z_L}{\psi_0}\;,
\end{equation}
where $\hat{S}^z_L(t)= e^{i\hat{H}^{(\mu)}t}\hat{S}^z_L e^{-i\hat{H}^{(\mu)}t} $. By selecting the initial product states in the computational basis, one can use a direct relationship $C^{(\mu)}(t) = s^{z}_L Q^{(\mu)}(t)$, where $s^z_L=\mel{\psi_0}{\hat{S}^z_L}{\psi_0}$, given that $\ket{\psi _0}$ is an eigenstate of $\hat{S}^z_L$. Hence, the short time values are given by $C_0 \equiv C^{(\mu)}(0)=1/4$ and $Q_0 \equiv Q^{(\mu)}(0)=\pm 1/2$. Subsequently, the autocorrelation functions are averaged over the Hamiltonian realizations as
\begin{equation}\label{eq:autocorrelation:av}
    C(t) \equiv  {\rm Avr} _\mu \{C^{(\mu)}(t)\}\;.
\end{equation} 
Such autocorrelators of local observables are also relevant due to their direct experimental accessibility~\cite{Choi16, Lukin19, Rispoli19}.
\begin{figure}[t]
    \centering
    \includegraphics[width=\columnwidth]{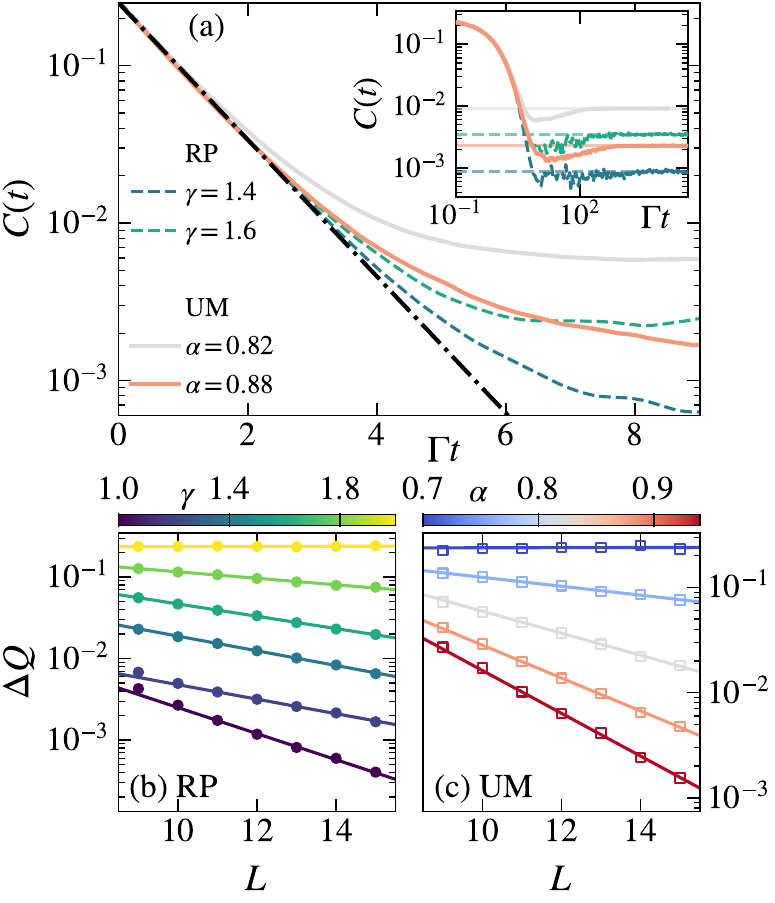}
    \caption
    { 
        (a) Averaged autocorrelation function $C(t)$ from Eq.~\eqref{eq:autocorrelation:av} at short times ($t\ll t_H$) for the UM [RP] model at $\alpha=0.82, 0.88$ [$\gamma=1.4, 1.6$] and ${L}=15$, marked with solid [dashed] lines.
        The time is rescaled by the Thouless energy $\Gamma$, numerically extracted from the spectral function as described in Sec.~\ref{sec:anomalous_dynamics}.
        The dash-dotted line shows a decay following $e^{-\Gamma t}$.
        Inset: Long-time dynamics of $C(t)$ as a function of the rescaled time $\Gamma t$. The horizontal solid [dashed] lines show the diagonal ensemble values, $C_{\rm DE} = {\rm Avr}_\mu\{C_{\rm DE}^{(\mu)}\}$, see also Eq.~\eqref{eq:autocorrelation:DE}, for the UM [RP] at the same values of $\alpha$ [$\gamma$].
        (b,c) Scaling of $\Delta Q$ as function of system size $L$ for the RP model [panel (b)] and UM model [panel (c)] for various values of $\gamma$ and $\alpha$ in the fading ergodicity regime, respectively.
        The solid lines in panels (b,c) show exponential fits of $a2^{-2L/\eta_Q}$ to the results.
    }\label{figM2}
\end{figure}

The main panel of Fig.~\ref{figM2}(a) shows the averaged autocorrelation functions $C(t)$ from Eq.~\eqref{eq:autocorrelation:av} for the RP [UM] model at two values of $\gamma$ [$\alpha$], matched (with small error margin) by the condition in Eq.~\eqref{eq:RP model_UM_relation}.
In order to show that the decay is indeed exponential with the rate $\Gamma$, we rescale the time $t$ to $t \rightarrow \Gamma t$.
With this we observe a collapse of the curves for both models at short times, with a clear exponential decay for almost two orders of magnitude.

\subsection{Long-time averages and thermalization} \label{sec:thermalization}

The long time limit of the autocorrelation functions in Eq.~\eqref{eq:autocorrelation} is given by  
\begin{equation}\label{eq:autocorrelation:DE}
    C_{\rm DE}^{(\mu)} = \lim_{t\to\infty}C^{(\mu)}(t)=s^z_L Q_{\rm DE}^{(\mu)}\;,
\end{equation}
where $Q_{\rm DE}^{(\mu)}$ is the diagonal ensemble (DE) prediction~\cite{rigol_dunjko_07},
\begin{equation}\label{eq:quench:DE}
    Q_{\rm DE}^{(\mu)} \equiv \lim _{t\to \infty} Q^{(\mu)}(t) = \sum_{n=1}^{\mathcal{D}}(S^z_L)_{nn}\abs{c_n^{(\mu)}}^2\;,
\end{equation}
for a given state $\ket{\psi _0}$ and Hamiltonian realization $\hat{H}^{(\mu)}$. Here, the coefficients $c_n^{(\mu)}=\braket{n^{(\mu)}}{\psi_0}$ represent the overlaps of the initial state $\ket{\psi _0}$ with the eigenstates of $\hat{H}^{(\mu)}$. 
The inset of Fig.~\ref{figM2}(a) shows the long-time dynamics of $C(t)$, and the horizontal lines show the DE predictions for the autocorrelation functions in Eq.~\eqref{eq:autocorrelation:DE}, averaged over Hamiltonian realizations $\mu$.

We also observe in the inset of Fig.~\ref{figM2}(a) that, for a given system size $L$ and the same decay rate $\Gamma$ in both models, the autocorrelation functions saturate to different values.
This is not unexpected as the long-time limit depends on the coefficients $\{c_n^{(\mu)}\}$, and the finite-size differences between the two models may arise due to distinct distributions of these coefficients.
Specifically, the eigenstate coefficients of the RP model follow a Lorentzian broadening, with a width $\Gamma$, around the energy of the state~\cite{bogomolny_sieber_18}.
On the other hand, for the UM model, it was found that the distribution of eigenstate coefficients follows a generalized hyperbolic distribution with a broad local density of states~\cite{bogomolny_sieber_18b}.
In Sec.~\ref{sec:fluctuations}, we study these differences using the inverse participation ratios.

Nevertheless, we argue that despite the different values of $Q_{\rm DE}^{(\mu)}$ in finite systems, the difference w.r.t.~the microcanonical ensemble (ME) prediction, $Q^{(\mu)}_{\rm ME}$, vanishes in the thermodynamic limit.
We define the ME prediction $Q_{\rm ME}^{(\mu)}$ as
\begin{equation}\label{eq:me}
    Q^{(\mu)}_{\rm ME} \equiv \frac{1}{\mathcal{N}_{\epsilon _i, \Delta _\epsilon}} \sum _{|E_n^{(\mu)} - \epsilon _{\psi_0} | < \Delta_{\epsilon}} (S_L^z)_{nn} \,.
\end{equation}
It is calculated at the initial state energy $\epsilon_{\psi_0}$ with a spread of $\Delta_\epsilon$. Thermalization is ensured when the DE result matches the ME predictions.
Therefore, we measure the deviations from the ME using
\begin{equation} \label{eq:deltaQ}
    \Delta {Q} = {\rm Avr}_\mu \left\{| {Q}_{\rm DE}^{(\mu)} - {Q}^{(\mu)} _{\rm ME} |\right\} \;.
\end{equation}
In Fig.~\ref{figM2}(b), we present the results of $\Delta Q$, see Eq.~\eqref{eq:deltaQ}, for the RP model, while Fig.~\ref{figM2}(c) illustrates the results for the UM model. For both models, we find good agreement of the two ensembles in the entire fading ergodicity regime with deviations that vanish in the thermodynamic limit.
We fit an exponential function $a2^{-2L/\eta_Q}$ to the results to show that it ultimately vanishes in the thermodynamic limit in the fading ergodicity regime below [above] the critical point $\gamma<\gamma_c$ [$\alpha>\alpha_c$] for the RP [UM] model.
Hence, the system reaches thermal equilibrium.

As a side remark, we note that the choice of the width of the ME window, $\Delta_\epsilon$, may not be straightforward.
In general, it should be chosen such that only a vanishing fraction of states contributes to ensemble averages.
In the UM model, where the eigenstates are either delocalized or weakly multifractal, we select $\Delta_\epsilon=0.05$, as altering the width of the window does not affect the outcome significantly.
On the other hand, in the RP model, the eigenstates are fractal~\cite{kravtsov_khaymovich2015}, making the selection of $\Delta_\epsilon$ rather non-trivial.
In Appendix~\ref{sec:app:ME} we find that choosing a window width of $\Delta_\epsilon\gg\Gamma$, i.e., the microcanonical window being larger than the width of the local density of states, leads to clear signatures of thermalization. For that reason, we consider $\Delta_\epsilon=1$ for all values of $\gamma$.


\subsection{Long-time fluctuations\label{sec:fluctuations}}

In this section, we study the fluctuations around the infinite time values of the autocorrelation functions, cf.~Eq.~\eqref{eq:autocorrelation:DE}. In the fading ergodicity regime of the quantum sun model, it was argued that the magnitude of fluctuations decays to zero in the thermodynamic limit~\cite{kliczkowski_vidmar2024}.
Here, we demonstrate that this is also true for the UM and RP models. 

We define the variance of the temporal fluctuation for the initial state $\ket{\psi _0}$ and the Hamiltonian realization $\mu$ as 
\begin{equation}\label{eq:temporal_state}
    [\sigma _t^{(\mu)}]^{2} = \frac{1}{N_t}\sum_{t=t_0}^{t_\infty}
    \qty(Q^{(\mu)}(t)-Q_{\rm DE}^{(\mu)})^2\;,
\end{equation}
where $N_t$ is the number of time steps in the interval $t\in[t_0,t_\infty]$, while $Q^{(\mu)}(t)$ is defined in Eq.~\eqref{eq:quench} and its DE prediction $Q_{\rm DE}^{(\mu)}$ is defined in Eq.~\eqref{eq:quench:DE}.
We set $t_0=t_H$ and $t_\infty = 10 t_H$, such that the signal is governed by the late-time dynamics; see also the discussion in Appendix~\ref{sec:app:shor_time_effect}.
Subsequently, the temporal fluctuations are averaged over Hamiltonian realizations as
\begin{equation}\label{eq:temporal}
    \sigma _t^2 \equiv  {\rm Avr} _\mu \{[\sigma _t^{(\mu)}]^{2}\}\;.
\end{equation}
The main question that we ask here is whether and how do the temporal fluctuations decay with the Hilbert space dimension ${\cal D}$.
We parametrize their decay with $\cal D$ as
\begin{equation} \label{def_sigmat_scaling}
    \sigma_t^2 \propto {\cal D}^{-2/\eta_t}\;,
\end{equation}
where the decay exponent $\eta_t$ may or may not be equal to the exponent $\eta$ of the fluctuations of the matrix elements introduced in Eq.~\eqref{eq:fading}.

It is known~\cite{dalessio_kafri_16} that there exists an initial-state independent upper bound on the temporal fluctuations given by the maximal off-diagonal matrix element,
$[\sigma _t^{(\mu)}]^{2}\leq\max_{\alpha\neq\beta}|O_{\alpha\beta}^{(\mu)}|^2$. 
If this bound was tight, it would imply $\eta_t = \eta$.

However, here we consider another, initial-state dependent upper bound that was derived in the context of typicality~\cite{reimann_2007, winter_linden_2009},
\begin{equation} \label{sigma_t_bound}
    [\sigma _t^{(\mu)}]^{2}\leq\norm{O}_1/\mathcal{D}^{(\mu)}_{\rm eff}\;,
\end{equation}
where $\mathcal{D}^{(\mu)}_{\rm eff}=1/(\sum_n |c^{(\mu)}_n|^4)$ is the effective dimension of the initial state in the Hamiltonian eigenbasis, $c_n^{(\mu)} =  \langle n^{(\mu)}|\psi_0\rangle$, and $\norm{O}_1$ is the operator norm defined as the difference between its maximal and minimal eigenvalues, $\norm{O}_1 = \max_i\{O_i\}-\min_i\{O_i\}$. 
Since the observable of interest has eigenvalues $\pm1/2$, it follows that $||S^z_L||_1=1$. 
Hence, the effective dimension $\mathcal{D}^{(\mu)}_{\rm eff}$ is given by the participation ratio of the initial state $|\psi_0\rangle$ in the eigenbasis of the Hamiltonian.
We express the corresponding IPR of $|\psi_0\rangle$ as 
\begin{equation} \label{def_ipr_0}
 {\rm IPR}^{(0)} = \sum_n |c^{(\mu)}_n|^4 = \sum_n |\langle n^{(\mu)}|\psi_0\rangle|^4 \equiv P_{s}^{(\mu)}\,,
\end{equation}
which should be contrasted with Eq.~\eqref{def_ipr_eig} that defines ${\rm IPR}^{\rm(eig)}$, i.e., the IPR of a Hamiltonian eigenstate in the computational basis.
\begin{figure}[t]
    \centering
    \includegraphics[width=\columnwidth]{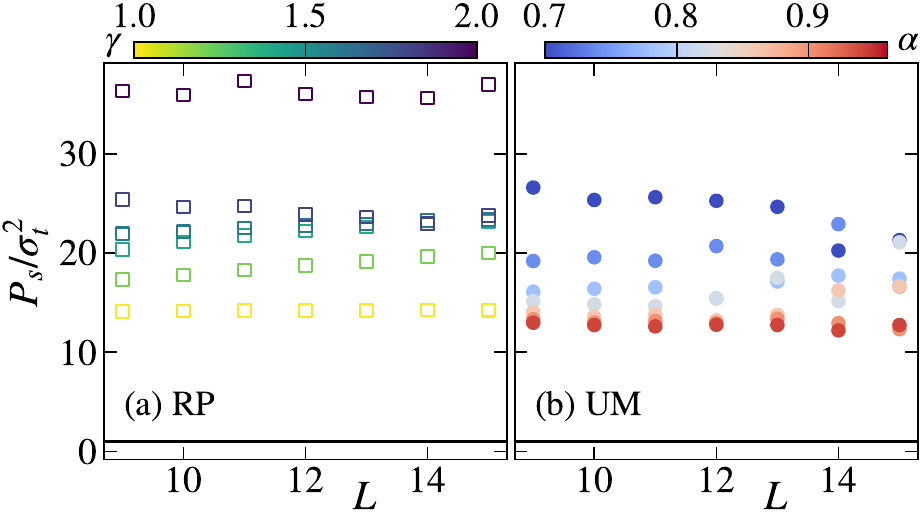}
    \caption
    { 
        Ratio $P_{s}/\sigma _t^2$ of the inverse participation ratio $P_s$ and the temporal fluctuations $\sigma_t^2$ versus the system size ${L}$ for the UM [RP] model at various $\alpha$ [$\gamma$].
        The solid vertical lines at the bottom show the lower bound $P_{s}/\sigma _t^2=1$ given by Eq.~\eqref{eq:bound}.
    }
    \label{figM3a}
\end{figure}
In Eq.~\eqref{def_ipr_0} we also denoted the IPR of $|\psi_0\rangle$ as $P_s^{(\mu)}$ to highlight that it is related to the infinite-time limit of the survival probability of the initial state $\ket{\psi _0}$.
We define the survival probability as
\begin{equation}\label{eq:survival}
    P^{(\mu)}(t)=\abs{\mel{\psi_0}{e^{-i\hat{H}^{(\mu)}t}}{\psi_0}}^2\;,
\end{equation}
and its infinite-time limit as $P_{s}^{(\mu)} = \lim_{t \rightarrow \infty} P^{(\mu)}(t)$. 
One can then express Eq.~\eqref{sigma_t_bound} as
$[\sigma _t^{(\mu)}]^{2}\leq P_{s}^{(\mu)}$ for this choice of an observable, and after averaging over the Hamiltonian realizations, one gets
\begin{equation}~\label{eq:bound}
    \sigma _t^2 \leq P_{s}^{} \;,
\end{equation}
where $P_{s} \equiv {\rm Avr} _\mu \{P_{s}^{(\mu)}\}$. 
We test Eq.~\eqref{eq:bound} in Fig.~\ref{figM3a}, in which we present the ratio $P_{s}/\sigma _t^2$ that should satisfy $P_{s}/\sigma _t^2 \geq 1$.
Indeed, we observe that Eq.~\eqref{eq:bound} is valid for all cases under consideration.

We further parametrize the scaling of $P_s$ with the Hilbert-space dimension as 
\begin{equation} \label{def_Ps_scaling}
    P_{s}^{} \propto \mathcal{D}^{-d_2^{(0)}}\,,
\end{equation}
where the fractal dimension of a single initial state, $d_2^{(0)}$, is expected to be similar, but not necessary equal, to the eigenstate fractal dimension $d_2^{({\rm eig})}$ from Eq.~\eqref{def_ipr_eig}.
Combining Eq.~\eqref{def_sigmat_scaling} with Eq.~\eqref{def_Ps_scaling}, one finds the following bound on the exponent $\eta_t$ of the temporal fluctuations,
\begin{equation}~\label{eq:bound2}
    2/\eta_t \geq d_2^{(0)} \;.
\end{equation}
The main question is whether Eq.~\eqref{eq:bound2} only represents a bound, or can the bound be saturated for the quantum quenches of interest.
In most cases in Fig.~\ref{figM3a}, the ratio $P_{s}/\sigma _t^2$ does show tendency towards saturation, which then suggests a saturation of the bound in Eq.~\eqref{eq:bound2}, i.e., $2/\eta_t \approx d_2^{(0)}$. 
We note that the numerical fits to extract the exponents $2/\eta_t$ and $d_2^{(0)}$ can give values that slightly violate Eq.~\eqref{eq:bound2}. However, this can be interpreted as finite size effects, as Eq.~\eqref{eq:bound} is still satisfied in such cases.

\begin{figure}[t]
    \centering
    \includegraphics[width=0.9\columnwidth]{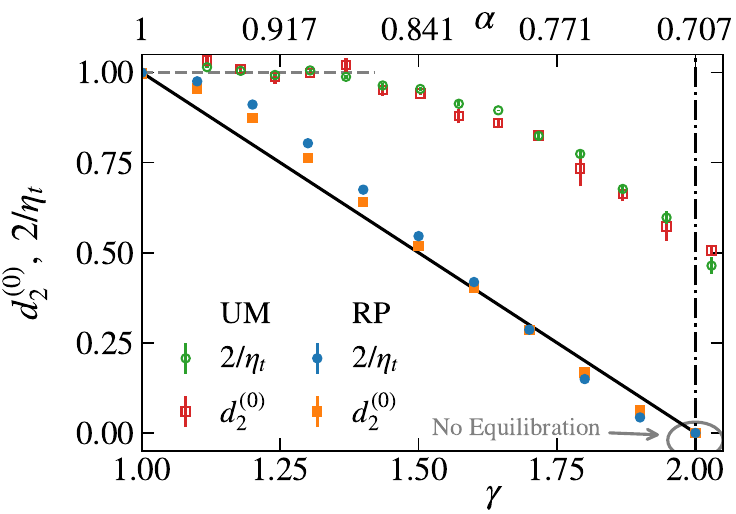}
    \caption
    { 
        Comparison of the exponent of the temporal fluctuations $2/\eta_t$ (circles) and the fractal dimension $d_2^{(0)}$ of the initial state $\ket{\psi_0}$ (squares) for the UM [RP] model, versus $\alpha$ [$\gamma$].
        We extract $2/\eta_t$ using Eq.~\eqref{def_sigmat_scaling} and $d_2^{(0)}$ using Eq.~\eqref{def_Ps_scaling}.
        Black solid line denotes the function $d_2^{\rm(eig)}=2-\gamma$ for the RP model.
        Vertical dash-dotted line shows the critical point for the ergodicity breaking transition, at which there is no equilibration in the RP model.
    }
    \label{figM3}
\end{figure}

In Fig.~\ref{figM3} we show the exponents $2/\eta_t$ and $d_2^{(0)}$ for the RP [UM] model versus $\gamma$ [$\alpha$].
The exponents were obtained by using Eqs.~\eqref{def_sigmat_scaling} and~\eqref{def_Ps_scaling} for the system sizes $8 < {L} \leq 16$. 
Remarkably, the two exponents agree well with each other for each of the two models, suggesting that it is reasonable to consider Eq.~\eqref{eq:bound2} as an equality rather than an inequality.

Nevertheless, the values of the exponents are rather different between the two models.
In the UM model, $d_2^{(0)}$ of the initial state $|\psi_0\rangle$ remains non-zero at the critical point $\alpha=\alpha _c=1/\sqrt{2}$, suggesting that the exponent $2/\eta_t$ remains finite.
The latter implies, on the one hand, equilibration of the observable after a quantum quench at the critical point.
On the other hand, finite $2/\eta_t$ suggests that $\eta_t$, for the initial state $|\psi_0\rangle$, does not diverge at the critical point, and hence it is different from the exponent $\eta$ from Eq.~\eqref{eq:fading} that diverges at the critical point.
Similar behavior was observed for the quantum sun model in Ref.~\cite{kliczkowski_vidmar2024}.

The situation is different in the RP model, for which $d_2^{(0)}$ of the initial state $|\psi_0\rangle$ vanishes at the critical point~$\gamma=\gamma_c=2$.
Vanishing of $d_2^{(0)}$ implies the vanishing of $2/\eta_t$, i.e., the lack of equilibration after a quantum quench at the critical point of the RP model.
Moreover, $d_2^{(0)}$ behaves similarly as $d_2^{\rm (eig)} = 2-\gamma$ that was introduced below Eq.~\eqref{def_ipr_eig}, see the solid line in Fig.~\ref{figM3}.
As a consequence, we find that the exponent $\eta_t$ that characterizes temporal fluctuations agrees with the exponent $\eta$ obtained from the fluctuations of matrix elements in Sec.~\ref{sec:anomalous_dynamics}.

\section{Spectral properties of the noise in the steady state} \label{sec:noise}

We now characterize the statistical properties of the temporal fluctuations beyond the variance studied in Sec.~\ref{sec:fluctuations}.
Defining the dynamics of the observable $\hat O = \hat S_L^z$ above its steady-state value as
\begin{equation}
    \delta Q^{(\mu)}(t) = \qty(Q^{(\mu)}(t)-Q^{(\mu)}_{\rm DE}) \;,
\end{equation}
we define the Fourier transform of $\delta Q^{(\mu)}(t)$ as
\begin{align}
 \mathcal{F}^{(\mu)}_Q(\omega)&=\int_{-\infty}^\infty dt e^{i\omega t} \delta Q^{(\mu)}(t) \nonumber\\
 &=\sum_{n\neq m}\qty(c_n^{(\mu)})^*c_m^{(\mu)} O_{nm}^{(\mu)}\delta(\omega-\omega_{nm})\;.
\end{align}
This allows us to define the power spectrum for a given Hamiltonian realization,
\begin{align}\label{app:eq:power_spec:mu}
    S^{(\mu)}(\omega) &= |\mathcal{F}^{(\mu)}_Q(\omega)|^2 \nonumber\\
    &= \sum_{n\neq m}\abs{c_n^{(\mu)}}^2\abs{c_m^{(\mu)}}^2\abs{O_{nm}^{(\mu)}}^2\delta(\omega-\omega_{nm})\;.
\end{align}
In the actual numerical calculations, we calculate the coarse-grained power-spectrum for a given Hamiltonian realization,
\begin{equation}\label{eq:power_spec_mu}
        \mathcal{S}^{(\mu)}(\omega)=\frac{\mathcal{D}}{\mathcal{M}'_\omega}\sum_{\substack{n\neq m:\\ |\omega_{nm}-\omega|<\delta\omega/2}}\abs{c_n^{(\mu)}}^2\abs{c_m^{(\mu)}}^2\abs{O_{nm}^{(\mu)}}^2\;,
\end{equation}
and we subsequently average the result over the Hamiltonian realizations,
\begin{equation}\label{eq:power_spec}
    \mathcal{S}(\omega)={\rm Avr}_\mu\{\mathcal{S}^{(\mu)}(\omega)\}\;.
\end{equation}
The factor $\mathcal{D}$ in Eq.~\eqref{eq:power_spec_mu} arises due to the coarse-graining.
Equation~\eqref{app:eq:power_spec:mu} suggests that the power spectrum consists of two contributions, i.e., the overlaps $c_n^{(\mu)}$ of the initial state $|\psi_0\rangle$ in the Hamiltonian eigenbasis, and the observable matrix elements $O_{nm}^{(\mu)}$.
Below we consider both contributions separately and explore to what extent can the power spectrum be decomposed into a product of these contributions.

\subsection{Decomposition of the power spectrum\label{sec:power}}

The first contribution to the power spectrum stems from the overlaps $c_n^{(\mu)}$.
They exhibit analogy with the Fourier transform of the survival probability $P^{(\mu)}(t)$ defined in Eq.~\eqref{eq:survival}. 
Subtracting from the later its steady-state value $P_s^{(\mu)}$, the Fourier transform reads
\begin{align}
    P^{(\mu)}(\omega) &= \int_{-\infty}^\infty dt e^{i\omega t} \left( P^{(\mu)}(t) - P_s^{(\mu)}\right) \nonumber\\
    &= \sum_{n\neq m}\abs{c_n^{(\mu)}}^2\abs{c_m^{(\mu)}}^2\delta(\omega-\omega_{nm})\,.
\end{align}
We refer to $P^{(\mu)}(\omega)$ as the spectrum of the local density of states (shortly, the spectrum of the LDoS), since it corresponds to the correlation function of the LDoS of the initial state $|\psi_0\rangle$, $\rho_{\rm LDoS}(E) = \sum_n |c_n^{(\mu)}|^2 \delta(E-E_n^{(\mu)})$~\cite{cuevas_kravtsov2007, Kravtsov10, khaymovich_kravtsov2021}.
Numerically, we express the spectrum of the LDoS as
\begin{equation}\label{eq:k_function:coarse}
    {\cal P}^{(\mu)}(\omega)=\frac{\mathcal{D}}{\mathcal{M}_\omega'}\sum_{\substack{n\neq m:\\ |\omega_{nm}-\omega|<\delta\omega/2}}\abs{c_n^{(\mu)}}^2\abs{c_m^{(\mu)}}^2 \,,
\end{equation}
where the coarse-graining is performed similarly to the one applied in Eq.~\eqref{eq:f_function:coarse:mu}.
Finally, we average over the Hamiltonian realizations to obtain
\begin{equation}
    {\cal P}(\omega)={\rm Avr}_\mu\qty{{\cal P}^{(\mu)}(\omega)} \,.
\end{equation}

\begin{figure}[t]
    \centering
    \includegraphics[width=\columnwidth]{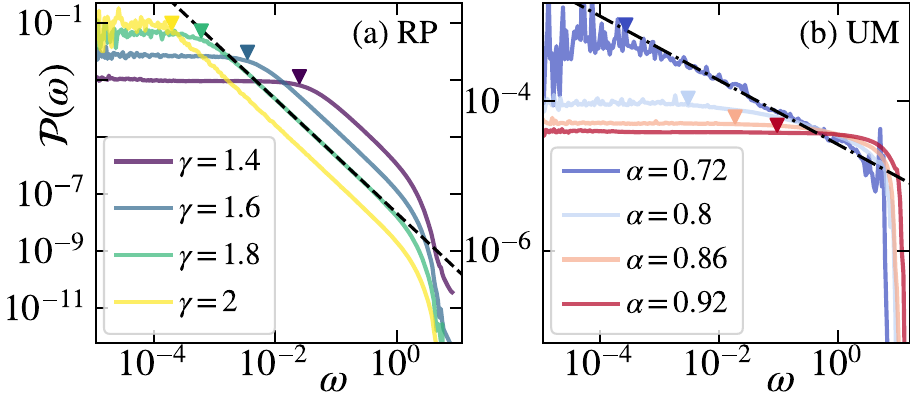}
    \caption{
    Coarse-grained spectrum of the LDoS, ${\cal P}(\omega)$, see Eq.~\eqref{eq:k_function:coarse}, versus $\omega$.
    (a) RP model, (b) UM model, both at ${L}=15$.
    Triangles denote the position of the Thouless energy, extracted from the spectral function $\mathcal{O}(\omega)$ using Eq.~\eqref{eq:integrated_spec}, see Fig.~\ref{fig:fig_O}.
    (a) Dashed line shows a Lorentzian decay, $\propto 1/\omega^2$.
    (b) Dash-dotted line is obtained using the Chalker ansatz~\cite{Chalker88,Chalker90}, $\propto 1/\omega^{1-d_2^{(0)}}$, where $d_2^{(0)}\approx0.57$ is the fractal dimension of the initial state $|\psi_0\rangle$ in the Hamiltonian eigenbasis at $\alpha=0.72$, i.e., in the vicinity of the critical point. 
    }
    \label{fig:fig_K}
\end{figure}

Examples of the spectra of LDoS, ${\cal P}(\omega)$, are shown in Figs.~\ref{fig:fig_K}(a) and~\ref{fig:fig_K}(b) for the RP model and the UM model, respectively. 
In the RP model, see Fig.~\ref{fig:fig_K}(a), we find a Lorentzian form of ${\cal P}(\omega)$ with a plateau at $\omega<\Gamma$, followed by a decay $\propto \omega^{-2}$ at larger frequencies~\cite{kravtsov_khaymovich2015}.
This shape of ${\cal P}(\omega)$ arises from the Lorentzian broadening of the eigenstate coefficients observed in the LDoS~\cite{kravtsov_khaymovich2015, bogomolny_sieber_18, khaymovich_kravtsov2021, skvortsov_kravtsov2022}, and it corresponds to an exponential decay in time of the survival probability~\cite{DeTomasi19}.

In contrast, in the UM model, long power-law tails emerge with an exponent different from $2$ as the ergodicity braking transition point is approached, see Fig.~\ref{fig:fig_K}(b). 
This corresponds to the power-law decay in time of the survival probability, which was previously observed at the ergodicity breaking transition point of the quantum sun model~\cite{hopjan_vidmar_23a} and the UM model~\cite{hopjan_vidmar_23b}.
The exponent of the power-law decay can be quantified by the Chalker's ansatz~\cite{Chalker88,Chalker90}, which states that at the critical point, ${\cal P}(\omega)$ decays as ${\cal P}(\omega) \propto \omega^{-1+d_2^{(0)}}$, where $d_2^{(0)}$ represents the fractal dimension of the initial state at the critical point. 
We test this ansatz in Fig.~\ref{fig:fig_K}(b) and find remarkable agreement at low frequencies.

\begin{figure}[!t]
    \centering
    \includegraphics[width=\columnwidth]{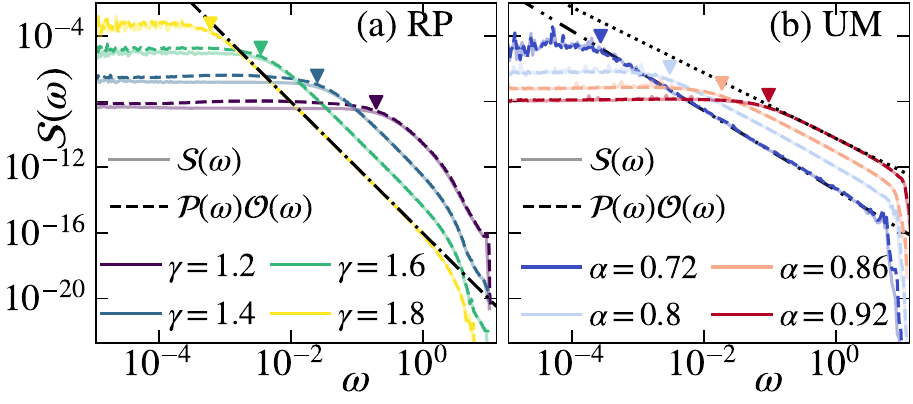} 
    \caption{
        Power spectrum $\mathcal{S}(\omega)$ from Eq.~\eqref{eq:power_spec}, see solid lines, compared to the decomposed power spectrum $\mathcal{S}^{'}(\omega)$ from Eq.~\eqref{eq:power_spec_conjg}, see dashed lines, for (a) RP model, and (b) UM model, both at ${L}=15$. 
        The triangles denote the Thouless energy $\Gamma$ extracted from the spectral function $\mathcal{O}(\omega)$ using Eq.~\eqref{eq:integrated_spec}, see Fig.~\ref{fig:fig_O}.
        (a) Dash-dotted line shows a $\propto \omega^{-4}$ decay, which is a consequence of the Lorentzian function of $\mathcal{O}(\omega)$, see Fig.~\ref{fig:fig_O}(a), as well as the Lorentzian function of ${\cal P}(\omega)$, see Fig.~\ref{fig:fig_K}(a). 
        (b) Dashed-double-dotted line is the product of the Lorentzian function for the spectral function $\mathcal{O}(\omega)$ from Fig.~\ref{fig:fig_O}(b) and the Chalker's ansatz~\cite{Chalker88,Chalker90} for ${\cal P}(\omega)$ in Fig.~\ref{fig:fig_K}(b), which gives rise to the function $\propto 1/\omega^{3-d_2^{(0)}}$.
        It describes well the results in the vicinity of the critical point of the UM model, e.g., at $\alpha=0.72$ when $d_2^{(0)}\approx0.57$.
        Dotted line shows a $\propto\omega^{-2}$ decay, since $d_2^{(0)}=1$ deep in the fading ergodicity regime of the UM model.
        }
    \label{figM5}
\end{figure}

The second contribution to the power spectrum is that of the matrix elements $O_{nm}^{(\mu)}$.
It is given by the spectral function $\mathcal{O}^{(\mu)} (\omega)$ introduced in Eq.~\eqref{eq:f_function:coarse:mu}.
This allows us to test the approximation for the power spectrum, which is decomposed into the product of two independent contributions,
\begin{equation}\label{eq:power_spec:decomposition}
    \begin{split}
          \mathcal{S}^{(\mu)'}(\omega) = {\cal P}^{(\mu)}(\omega)\mathcal{O}^{(\mu)} (\omega)\,.
    \end{split}
\end{equation}
Moreover, we average both contributions over the Hamiltonian realizations to obtain the approximation of the averaged power spectrum,
\begin{equation}\label{eq:power_spec_conjg}
    \mathcal{S}^{'}(\omega) \approx {\cal P}_{}(\omega)\mathcal{O}_{} (\omega) \;,
\end{equation}
which should be tested against the exact power spectrum from Eq.~\eqref{eq:power_spec}.

In Fig.~\ref{figM5} we compare the exact power spectrum $\mathcal{S}(\omega)$ from Eq.~\eqref{eq:power_spec} [solid lines] with the the approximation $\mathcal{S}^{'}(\omega)$ from Eq.~\eqref{eq:power_spec_conjg} [dashed lines].
They exhibit remarkable agreement in the entire fading ergodicity regime.
They suggest that the projections of the initial state can be, to a good approximation, considered as statistically independent from the matrix elements $O_{nm}^{(\mu)}$. 

As a consequence, the power spectrum can be used to obtain information about the Thouless energy for the given observable.
In Fig.~\ref{figM5}, the inverted triangles mark the Thouless energy $\Gamma$ extracted from the spectral function $\mathcal{O}(\omega)$ using Eq.~\eqref{eq:integrated_spec}, see Fig.~\ref{fig:fig_O}, which emerge very close to the onset of the plateau at small $\omega$ in the power spectrum.
One may also use the Chalker's ansatz to describe the power-law decay of the power spectrum at large $\omega$.
As an example, we show that the power-spectrum in the RP model close to the critical point, at $\alpha=0.72$, is well approximated by the function $\propto\omega^{-3+d_2^{(0)}}$, see Fig.~\ref{figM5}(b).
This is a consequence of the Lorentzian decay of the spectral function, $\mathcal{O}(\omega)$, and the power-law decay, $\propto\omega^{-1+d_2^{(0)}}$, of the spectrum of the LDoS, ${\cal P}(\omega)$.
On the other hand, in the RP model, see Fig.~\ref{figM5}(a), the power spectrum is well approximated by the function $\propto \omega^{-4}$, i.e., by the product of two Lorentzian functions.

\begin{figure}[t]
    \centering
    \includegraphics[width=\columnwidth]{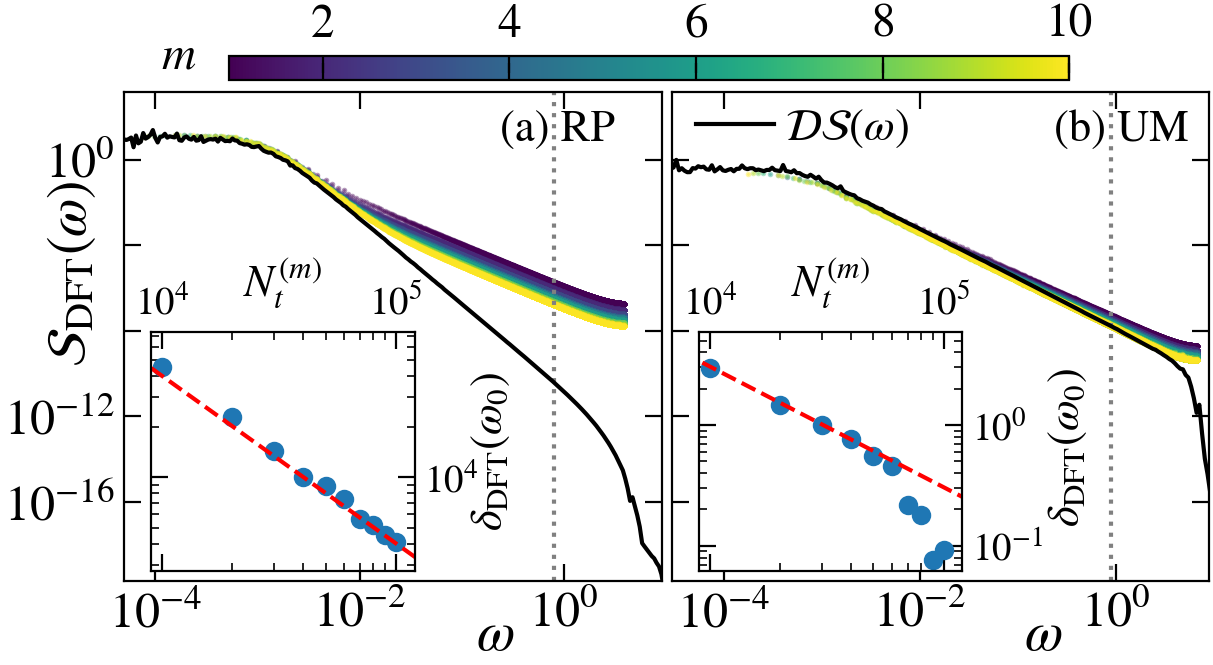}
    \caption
    {
        Power-spectrum obtained using the DFT, $\mathcal{S}_{\rm DFT}(\omega_k)$, see Eq.~\eqref{def_Sfft}, versus the power-spectrum $\mathcal{S}(\omega)$ from Eq.~\eqref{eq:power_spec}.
        Different colors for $\mathcal{S}_{\rm DFT}(\omega_k)$, from dark to bright, denote different number of time steps $N_t^{(m)}$, where the index $m$ is defined in Eq.~\eqref{def_timestep}.
        (a) RP model at $\gamma=1.8$ and (b) UM model at $\alpha=0.74$, for ${L}=14$.
        Insets: Scaling of the relative deviation $\delta_{\rm DFT}(\omega_0)$, see Eq.~\eqref{eq:power_spec:error}, versus $N_t^{(m)}$.
        It is measured at the frequency $\omega_0=\Delta E/10$, which is denoted by the vertical dotted lines in the main panel.
        Dashed red lines are fits of the function $a_0/N_t^{(m)}$ to the results, with $a_0\approx 3.98$ for the RP model and $a_0\approx5.2\cdot10^4$ for the UM model.
    }\label{figM6}
\end{figure}

\subsection{Discrete time approximation}

We now ask a question whether the features of the coarse-grained power spectrum can already be captured from a discrete series of $N_t$ stroboscopic time steps within the time interval $t\in\qty[t_H,t_H+N_t\delta t]$. To this end, we consider a discrete Fourier transform (DFT) of the subtracted observable,
\begin{equation}\label{app:eq:fourier:quench:discrete}
    \mathcal{F}^{(\mu)}_Q(\omega_k)= \frac{1}{\sqrt{N_t}} \sum_{n=0}^{N_t} e^{i\omega_k t_n}\qty(Q^{(\mu)}(t_n)-Q^{(\mu)}_{\rm DE})\;,
\end{equation}
where $t_n=t_H+n\delta t$ are the discrete times and the angular frequencies are $\omega_k=2\pi k/(N_t\delta t)$, with $k=-N_t,..,N_t$.
Thus, the maximal frequency that can be resolved is $2\pi/\delta t$. We choose $N_t=10^5$ and $\delta t=2\pi/\Delta E$.
The corresponding power spectrum is proportional to the squared modulus of the Fourier amplitudes,
\begin{equation}\label{eq:power_spectrum:fft}
    \mathcal{S}^{(\mu)}_{\rm DFT}(\omega_k)=(\delta t)^2|{\mathcal{F}_Q^{(\mu)}(\omega_k)}|^2\;,
\end{equation}
and we subsequently average the power spectrum over independent Hamiltonian realizations to obtain
\begin{equation} \label{def_Sfft}
    \mathcal{S}_{\rm DFT}(\omega) = {\rm Avr}_\mu\{ \mathcal{S}^{(\mu)}_{\rm DFT}(\omega) \}\,.
\end{equation}

Comparison between the power spectrum $\mathcal{S}(\omega)$ from Eq.~\eqref{eq:power_spec} and $\mathcal{S}_{\rm DFT}(\omega)$ from Eq.~\eqref{def_Sfft} is shown in Fig.~\ref{figM6}.
The agreement is perfect at low frequencies, while deviations are observed at high frequencies.
However, we also observe that the high-frequency tails get progressively more accurate when the number of time steps increases.
To illustrate this, we consider in Fig.~\ref{figM6} a sequence of DFTs with an increasing number of time steps.
Specifically, for each integer $m=1,\dots,M$, where $M=10$, and fixed $N_t=10^5$, we define the effective number of time steps as
\begin{equation} \label{def_timestep}
    N_t \rightarrow N_t^{(m)} = \lfloor \frac{m}{M} N_t\rfloor\;,
\end{equation}
which sets the duration of the total time interval entering the DFT. 
In Fig.~\ref{figM6} we observe that enlarging $m$, at large frequency, the power spectrum computed from the DFT approaches the power spectrum from Eq.~\eqref{eq:power_spec}.
We also observe that the deviations between the two are larger for the RP model, see Fig.~\ref{figM6}(a), than for the UM model, see Fig.~\ref{figM6}(b).
We quantify the deviation by defining the relative deviation,
\begin{equation}\label{eq:power_spec:error}
    \delta_{\rm DFT}(\omega_0)=\abs{1-\frac{\mathcal{S}_{\rm DFT}(\omega_0)}{\mathcal{S}(\omega_0)}}\,,
\end{equation}
at a particular frequency $\omega_0=\Delta E/10$ in the large frequency region, where $\Delta E$ is the bandwidth defined in Appendix~\ref{sec:app:variance}.
We plot $\delta_{\rm DFT}(\omega_0)$ in Eq.~\eqref{eq:power_spec:error} for both models in the insets of Figs.~\ref{figM6}(a) and~\ref{figM6}(b).
In both cases, we find that $\delta_{\rm DFT}(\omega_0)$ vanishes with increasing $N_t^{(m)}$ as $a_0/N_t^{(m)}$ or faster, with the prefactor $a_0$ that is larger for the RP model than for the UM model.


\section{Conclusions} \label{sec:conclusion}

We showed that the structured random matrix models, such as the Rosenzweig-Porter model and the ultrametric model, when embedded in a many-body Hilbert space of spins-1/2, exhibit key signatures of fading ergodicity.
These features were previously identified in models with few-body interactions, such as the quantum sun model~\cite{kliczkowski_vidmar2024}.
Fading ergodicity is a precursor regime to the ergodicity breaking transition, in which short-range spectral statistics flow towards the GOE limit.
However, the associated Thouless time, which can be quantitatively estimated using the Fermi golden rule, rapidly approaches the Heisenberg time.
This behavior is manifested in the reduced fluctuations of the matrix elements of observables.

We analyzed fading ergodicity through quantum quench dynamics from initial product states in the computational basis. 
We focused on the dynamics of observables with the longest relaxation times, set by the Thouless time.
Hence, in the fading ergodicity regime, these observables equilibrate on timescales shorter than the Heisenberg time, and their long-time expectation values approach the microcanonical ensemble predictions, indicating thermalization.

That said, a detailed analysis of the quantum dynamics further reveals certain differences between the Rosenzweig–Porter model and the ultrametric model.
These differences manifest, for instance, in the variances of late-time temporal fluctuations, the survival probabilities of the initial product states, and the power spectra of observable fluctuations. 
Many of these features can be understood as dynamical signatures of previously established properties of the two models. 
In particular, a key role is played by the structure of Hamiltonian eigenstates in the computational basis, which ranges from fractal states in the Rosenzweig–Porter model~\cite{kravtsov_khaymovich2015} to fully delocalized, or weakly multifractal, states in the ultrametric model~\cite{suntajs_hopjan_24}. 
Nevertheless, despite the differences in the fractal dimensions of the eigenstates, fading ergodicity provides a unifying framework for understanding ergodicity breaking in both models.

We also highlight several additional insights that emerge from our study. 
First, we identified an approximate decomposition of the power spectrum into a product of the observable spectral function (i.e., the Fourier transform of its autocorrelation function) and the correlation function of the local density of states (i.e., the Fourier transform of the survival probability). 
In models such as the ultrametric model, the latter exhibits a power-law decay at the critical point, with an exponent related to the fractal dimension of the initial state~\cite{hopjan_vidmar_23b}. Consequently, the power spectrum itself encodes information about the fractal dimension of the initial state. 
Moreover, we showed that the power spectrum may also contain signatures of the Thouless energy. 
Finally, we investigated the differences between the exact power spectrum and its approximation arising from a finite number of time steps in the discrete Fourier transform of the observable dynamics, and we demonstrated the convergence of the approximate spectrum to the exact result.

\acknowledgements 
We acknowledge discussions with I. M. Khaymovich and P. \L yd\.{z}ba, and support from the Slovenian Research and Innovation Agency (ARIS), Research core funding Grants No.~P1-0044, N1-0273, J1-50005 and N1-0369, as well as the Consolidator Grant Boundary-101126364 of the European Research Council (ERC). M. H. acknowledges support from the Polish National Agency for Academic Exchange (NAWA)’s Ulam Programme (project BNI/ULM/2024/1/00124).
We gratefully acknowledge the High Performance Computing Research Infrastructure Eastern Region (HCP RIVR) consortium~\cite{vega1} and European High Performance Computing Joint Undertaking (EuroHPC JU)~\cite{vega2} for funding this research by providing computing resources of the HPC system Vega at the Institute of Information sciences~\cite{vega3}.
We further acknowledge resources for numerical calculations made available by the Wrocław Centre for Networking and Supercomputing.

\bibliographystyle{biblev1}
\bibliography{references}

\appendix

\section{Energy variance and bandwidth\label{sec:app:variance}} 

In the main text, we used the energy variance $\sigma_E^2$ and the energy bandwidth $\Delta E$ as measures of energy.
We define the energy variance as the variance of the density of states, $\smash{\sigma_E^2 = \Tr \{\hat{H} ^2 \} / \mathcal{D}}$, and the energy bandwidth as the difference between the maximal and minimal energy, $\Delta E = E_{\rm max}-E_{\rm min}$.

In the RP model, the energy variance scales as $\sigma_E = \sqrt{1+\mathcal{D}^{1-\gamma}}$. 
At $\gamma<1$, the density of states follows a Wigner semicircle, and hence the energy bandwidth is $\Delta E=4\sigma_E$. 
In this regime, the Thouless energy scales as $\Gamma\sim\Delta E$~\cite{venturelli_tarzia_23}.
At $\gamma>2$, the density of states is Gaussian, since it is dominated by $\hat{H}_0$.
Hence, the spectral extrema can be approximated as $\smash{E_{\rm max/min}\approx E_{\rm av}\pm\sigma_E\sqrt{2\ln{\mathcal{D} }}}$ through the extreme value statistics of a Gaussian distribution, where $E_{\rm av} \equiv \Tr \{ \hat{H}\}/\mathcal{D}$. Hence, it follows that the bandwidth can be estimated as $\Delta E=2\sigma_E\sqrt{2\ln{\mathcal{D} }}$.

In the UM model, the energy variance can be derived as
\begin{equation}
    \sigma_E^2 = \left(1 + \sum _{k=1}^{L} \alpha ^{2k} \right) = \left(1 + \frac{\alpha ^2 (1 - \alpha ^{2L})}{1 - \alpha ^2} \right)\;,
\end{equation}
where we have used the property that the normalized matrices $\hat{H}_k$ from Eq.~\eqref{eq:um_def} are mutually uncorrelated random matrices, with a unit variance $\Tr {\hat{H}_k^2}/\mathcal{D}=1$, such that the final trace reduces to a geometric series in $\alpha^{2L}$. Since the density of states in the UM model is Gaussian, similarly to the RP model at $\gamma \gg 1$, the bandwidth can be approximated accordingly as $\smash{E_{\rm max/min}\approx E_{\rm av}\pm\sigma_E\sqrt{2\ln{\mathcal{D} }}}$.

\section{Fading ergodicity and the diagonal matrix elements\label{sec:app:fading:diagonal}} 

In Sec.~\ref{sec:anomalous_dynamics}, we studied fluctuations of the low-frequency off-diagonal matrix elements in the fading ergodicity regime.
Here, we compare this result by studying the fluctuations of the diagonal matrix element of the same observable, $\hat O = \hat{S}^z_L$.
\begin{figure}[t]
    \centering
    \includegraphics[width=\columnwidth]{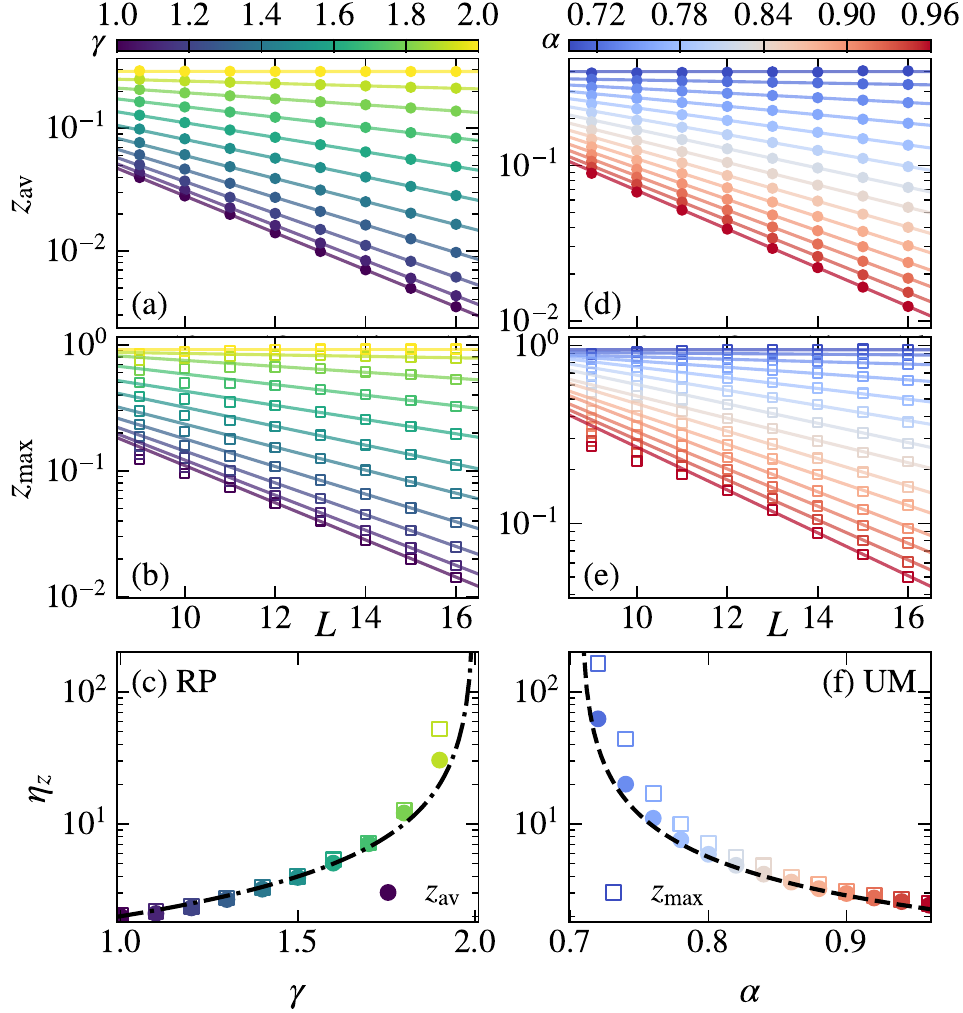}
    \caption
    {
    Fluctuations of the diagonal matrix elements in (a,b,c) the RP model, and (b,d,f) the UM model
    (a,d) Average eigenstate-to-eigenstate fluctuations $z_{\rm av}$, see Eq.~\eqref{eq:z_av}, versus ${L}$. 
    (b,e) The maximal outlier $z_{\rm max}$, see Eq.~\eqref{eq:z_max}, versus $L$.
    We fit the function $a\mathcal{D}^{-2/\eta_z}$ to the results in panels (a,b,c,d) for ${L}\geq 12$ and show the resulting values of $\eta_z$ in panels (c,f). 
    The dashed curves in (c,f) show the analytical prediction from Eq.~\eqref{eq:eta:prediction}, i.e., $\eta=2/(2-\gamma)$ [dash-dotted] for the RP model $\eta=2(1-\ln{\alpha}/\ln{\alpha_c})^{-1}$ [dashed] for the UM model.
    }\label{figA2}
\end{figure}

Specifically, we study the eigenstate-to-eigenstate fluctuations,
\begin{equation}\label{eq:z_n}
    z_n=\abs{O_{n+1,n+1}-O_{nn}}\;,
\end{equation}
and calculate the corresponding averages,
\begin{equation}\label{eq:z_av}
    z_{\rm av}={\rm Avr}_\mu\qty{{\rm Avr}_n\qty{z_n}} \,.
\end{equation}
They are averaged, for a given Hamiltonian realization, over $\min(500,\mathcal{D}/10)$ Hamiltonian eigenstates $\ket{n}$, and then over the Hamiltonian realizations $\hat{H}^{(\mu)}$.
We also consider the maximal outliers of a given Hamiltonian realization $\hat{H}^{(\mu)}$, averaged over the realizations, 
\begin{equation}\label{eq:z_max}
    z_{\rm max}={\rm Avr}_\mu\qty{\max_n\qty{z_n}}\;.
\end{equation}

For the RP model, we show the results for $z_{\rm av}$ and $z_{\rm max}$ in Figs.~\ref{figA2}(a) and~\ref{figA2}(b), respectively.
We observe the softening of fluctuations as one approaches the ergodicity breaking transition, which is a hallmark of fading ergodicity.
This result it similar to the scaling of the low-frequency off-diagonal matrix elements shown in Fig.~\ref{figM_eta}(a) of the main text.
We apply the fit $a\mathcal{D}^{-2/\eta_z}$ to the results and display the resulting values of $\eta_z$ for both $z_{\rm av}$ and $z_{\rm max}$ in Fig.~\ref{figA2}(c).
We find remarkable agreement with the prediction in Eq.~\eqref{eq:eta:prediction}, suggesting that $\eta_z$ indeed corresponds to the exponent $\eta$ introduced in Eq.~\eqref{eq:fading} in the main text.

We carry out the same analysis for the UM model, and we observe similar behavior, i.e., the softening of fluctuations in both $z_{\rm av}$, see Fig.~\ref{figA2}(d), and $z_{\rm max}$, see Fig.~\ref{figA2}(e), which is a hallmark of fading ergodicity.
We apply the fit $a\mathcal{D}^{-2/\eta_z}$ to the results in Figs.~\ref{figA2}(d) and~\ref{figA2}(e), and we plot the resulting exponent $\eta_z$ in Fig.~\ref{figA2}(c). 
Results for $\eta_z$ again show remarkable agreement with the analytical prediction from Eq.~\eqref{eq:eta:prediction}.


\section{Fading ergodicity and the level spacing ratio} \label{sec:gap_ratio}
\begin{figure}[t]
    \centering
    \includegraphics[width=0.9\columnwidth]{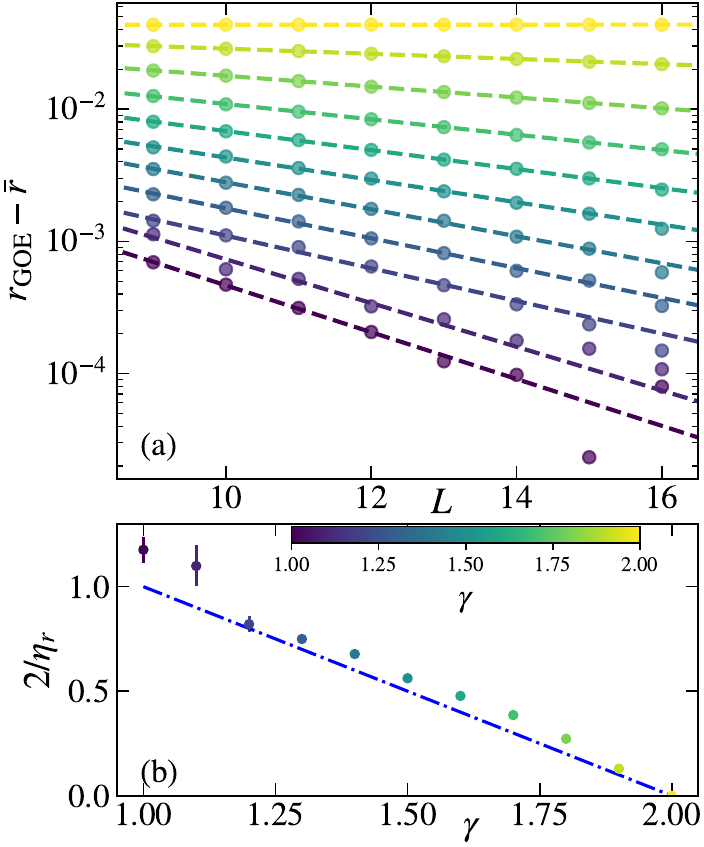}
    \caption
    {
        (a) Difference $r_{\rm GOE}-\bar r$ between the average level spacing ratio, $\bar r$, and the GOE prediction $r_{\rm GOE}$, in the RP model. 
        The colors correspond to different values of $\gamma$ as indicated in the legend in panel (b).
        The dashed lines represent a fit of the function $\propto\mathcal{D}^{-2/\eta _r}$ to the numerical results. 
        (b) Symbols denote the exponent $2/\eta_r$, extracted from the fits in panel (a), versus $\gamma$.
        The dashed-dotted line is the analytical prediction for the fluctuations of matrix elements from Eq.~\eqref{eq:eta:prediction}, $2/\eta=2-\gamma$.
    }\label{figA3}
\end{figure}

In this section, we show that, in the fading ergodicity regime, the short-range spectral statistics flow towards the GOE limit, however, they do approach this limit with a different rate.
Thus, fading ergodicity can also be detected from the analysis of standard spectral indicators such as the average level spacing ratio. We define the level spacing ratio $r_n^{(\mu)}$ for an eigenstate $n$ and Hamiltonian realization $\mu$ as
\begin{equation}\label{eq:gap_ratio}
    r_n^{(\mu)}=\frac{\min\qty{s_n^{(\mu)},s_{n+1}^{(\mu)}}}{\max\qty{s_n^{(\mu)},s_{n+1}^{(\mu)}}}\;,
\end{equation}
where $s_n^{(\mu)}=E_{n+1}^{(\mu)}-E_n^{(\mu)}$ are the consecutive level spacings.
The average level spacing ratio $\bar r$ is then obtained by averaging $r_n^{(\mu)}$ over the entire spectrum for a given Hamiltonian realization, and then over different Hamiltonian realizations,
\begin{equation}\label{eq:r_av}
    \bar r={\rm Avr}_\mu\qty{{\rm Avr}_n\qty{r_n^{(\mu)}}} \,.
\end{equation}
The GOE prediction is $r_{\rm GOE}=0.5307$~\cite{Atas_13}, while in the systems with Poisson statistics one gets $r_{\rm P}=2\ln{2}-1$~\cite{Oganesyan07, Giraud2018}.
Although we expect $r$ to comply with the GOE prediction in the fading ergodicity regime~\cite{kliczkowski_vidmar2024}, we show below that the corrections to the GOE prediction carry information about proximity to the ergodicity breaking transition.

Using the RP model as an example, we plot in Fig.~\ref{figA3}(a) the difference $r_{\rm GOE} - \bar r$ versus $L$ in the fading ergodicity regime.
Remarkably, we find that the differences decay exponentially with $L$, with a rate that decreases as one approaches the ergodicity breaking transition.
This behavior is very similar to the one observed in the decay of the matrix elements fluctuations, see, e.g., Fig.~\ref{figM_eta}(a) for the off-diagonal matrix elements and Figs.~\ref{figA2}(a) and~\ref{figA2}(b) for the diagonal matrix elements.

We fit in Fig.~\ref{figA3}(a) the results for a fixed $\gamma$ with a function $\propto\mathcal{D}^{-2/\eta_r}$, and we plot the extracted values $2/\eta_r$ versus $\gamma$ in Fig.~\ref{figA3}(b).
The results are very close to the analytical prediction derived for the fluctuations of matrix elements, $2/\eta = 2-\gamma$, see Eq.~\eqref{eq:eta:prediction}.
Hence, one observes $\eta_r \approx \eta$, which hints at the universal manifestation of fading ergodicity in different properties of finite quantum systems.

\section{Width of microcanonical ensemble\label{sec:app:ME}}

For the majority of quantum systems, the choice of the width of the microcanonical ensemble (ME) energy window does not considerably impact the value of ME prediction, provided that the window contains enough states but remains sufficiently narrow in energy.
However, this is not always guaranteed for systems with fractal states. 
In this case, the width of the local density of states may significantly affect the agreement between ME and DE predictions. 

\begin{figure}[!t]
    \centering
    \includegraphics[width=\columnwidth]{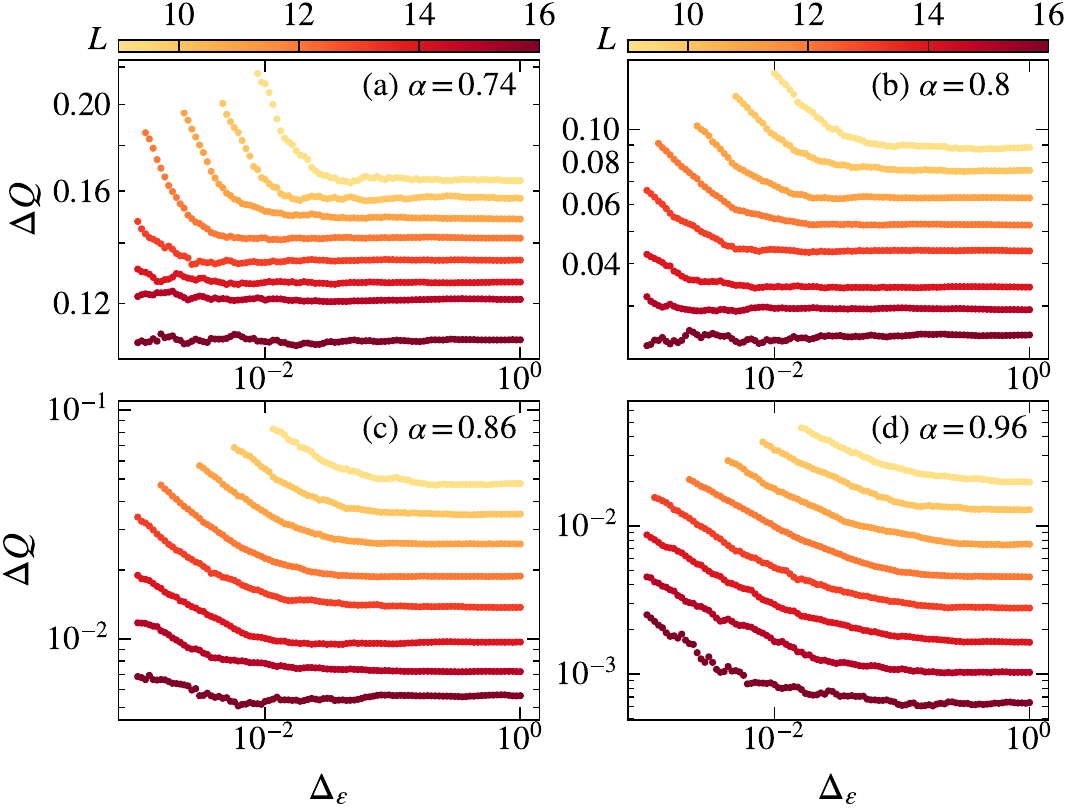}
    \caption
    {
        Dependence of $\Delta Q$, see Eq.~\eqref{eq:deltaQ}, on the width of the ME window $\Delta_\epsilon$, see Eq.~\eqref{eq:me}, in the UM model at different system sizes $L$.
        (a) $\alpha=0.74$, (b) $\alpha=0.80$, (c) $\alpha=0.86$, and (d) $\alpha=0.96$.
    }\label{figA4b}
\end{figure}

First, we consider the UM model with completely delocalized or weakly multifractal eigenstates.
In Fig.~\ref{figA4b} we show results for the differences between the ME and DE predictions $\Delta Q$, defined in Eq.~\eqref{eq:deltaQ}, as a function of the width of the ME window, $\Delta_\epsilon$, for various values of $\alpha$.
We find that both, in the vicinity of the ergodicity breaking transition and far away from it, the difference between DE and ME predictions is quite insensitive to the choice of the width of the ME window.
Therefore, we set $\Delta_\epsilon=0.05$ in the main text to study thermalization after a quantum quench in the UM model.

\begin{figure}[!t]
    \centering
    \includegraphics[width=\columnwidth]{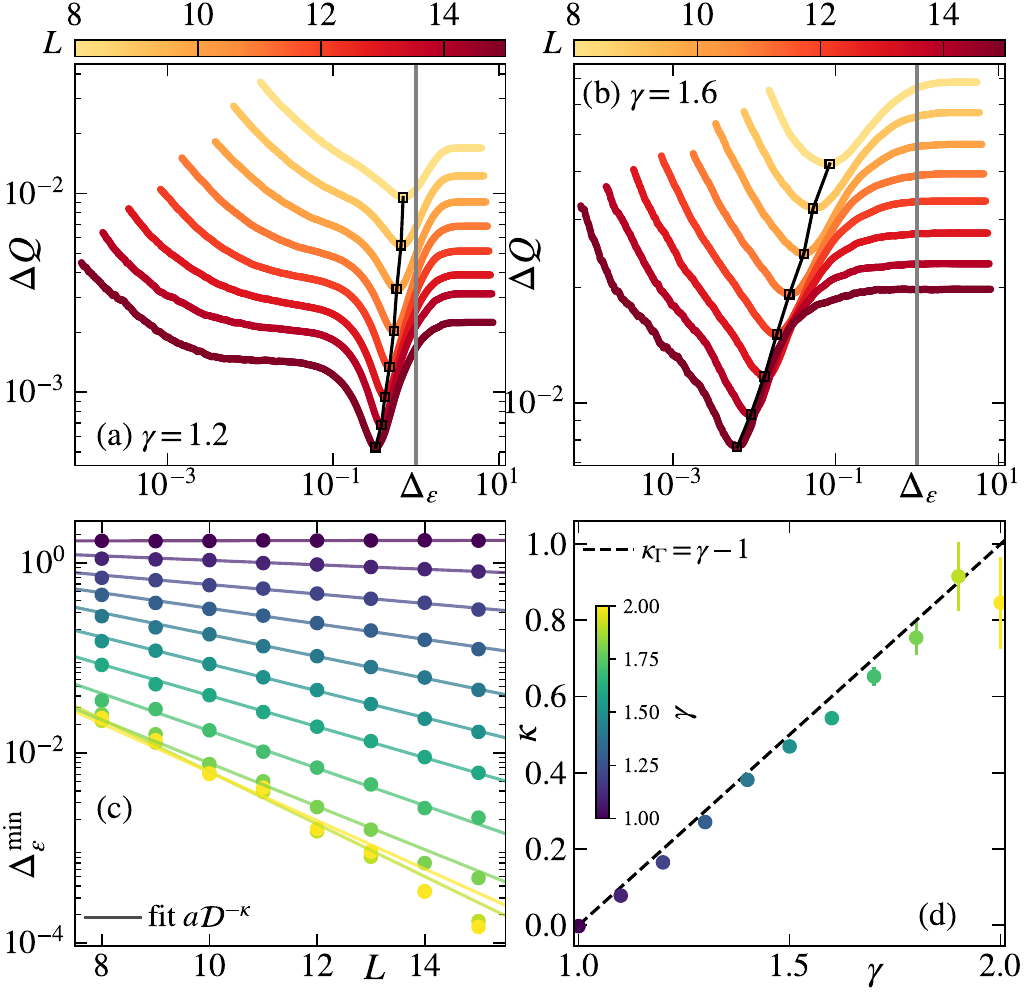}
    \caption
    {
        Dependence of $\Delta Q$, see Eq.~\eqref{eq:deltaQ}, on the width of the ME window $\Delta_\epsilon$, see Eq.~\eqref{eq:me}, in the RP model at different system sizes $L$.
        (a) $\gamma=1.1$ and (b) $\gamma=1.5$.
        In panels (a) and (b) the symbols mark the minimal values of $\Delta Q$.
        The vertical lines show $\Delta_\epsilon=1$, which is chosen in the main text.
        (c) Position of the minima of $\Delta Q$ as a function of system size $L$.
        We fit the function $\propto \mathcal{D}^{-\kappa}$ (solid lines) to the numerical results (symbols).
        (d) The exponent $\kappa$, extracted from the fits in panel (c), as a function of the model parameter $\gamma$.
        Results for $\kappa$ (symbols) are compared to the scaling of the Thouless energy $\Gamma$, see Eq.~\eqref{eq:RP:thouless}, which yields $\kappa_\Gamma=\gamma-1$ (dashed line). 
    }\label{figA4}
\end{figure}

We next study the effect of the width of the ME window, $\Delta_\epsilon$, on $\Delta Q$ in the RP model. 
In this case, the initial state of the quantum quench is fractal in the eigenbasis of the RP model, see Fig.~\ref{figM3} in Sec.~\ref{sec:quench}.
In Figs.~\ref{figA4}(a) and~\ref{figA4}(b) we show $\Delta Q$ as a function of $\Delta_\epsilon$ at $\gamma=1.2$ and $\gamma=1.6$.
Although the results exhibit a general trend for $\Delta Q$ to vanish exponentially with system size $L$, we also observe a minimum of $\Delta Q$ at certain values of $\Delta_\epsilon$.
As a consequence, the curves for $\Delta Q$ versus $\Delta_\epsilon$ may overlap for certain system sizes $L$, see, e.g., the results for $\Delta Q$ in range $\Delta_\epsilon\in\qty[10^{-2}, 10^{-1}]$ in Fig.~\ref{figA4}(b), which may yield a false impression of the absence of thermalization.

We extract $\Delta_\epsilon^{\rm min}$, i.e., the position of the minimum of $\Delta Q$, and plot it in Fig.~\ref{figA4}(c) versus the system size $L$.
Interestingly, $\Delta_\epsilon^{\rm min}$ scales exponentially towards zero, suggesting a connection to the width of the local density of states, and hence the Thouless energy $\Gamma$.
We fit the decay of $\Delta_\epsilon^{\rm min}$ by the function $a\mathcal{D}^{-\kappa}$, and we plot the extracted exponent $\kappa$ versus the model parameter $\gamma$ in Fig.~\ref{figA4}(d).
We compare the results to the corresponding exponent $\kappa_\Gamma = \gamma-1$, which is obtained from the scaling of the Thouless energy in Eq.~\eqref{eq:RP:thouless}, i.e., $\Gamma\propto\mathcal{D}^{1-\gamma}\equiv\mathcal{D}^{-\kappa_\Gamma}$.
We find that the numerically obtained values of $\kappa$ are indeed very close to $\kappa_\Gamma$, indicating a direct correlation of the minimum in $\Delta Q$ with the the Thouless energy and the width of the local density of states.

The nontrivial dependence of $\Delta Q$ on $\Delta_\epsilon$ in Fig.~\ref{figA4} shows the subtlety of choosing an appropriate ME energy window when considering initial states that are fractal in the Hamiltonian eigenbasis.
In order to keep the window small, yet larger than the Thouless energy, $\Delta_\epsilon\gg\Gamma$, we consider $\Delta_\epsilon=1$ in the main text for all values of $\gamma$.
We note that for values $\gamma\to 1$ the Thouless energy becomes $O(1)$, and hence $\Delta_\epsilon\sim\Gamma$.
In this case, the choice $\Delta_\epsilon=1$ is still reasonable since there is no drift of the minimum of $\Delta Q$ and hence there is no spurious signature of a non-decaying value of $\Delta Q$.

\section{Choice of time window to obtain variances of temporal fluctuations} \label{sec:app:shor_time_effect}

\begin{figure}[!t]
    \centering
    \includegraphics[width=\columnwidth]{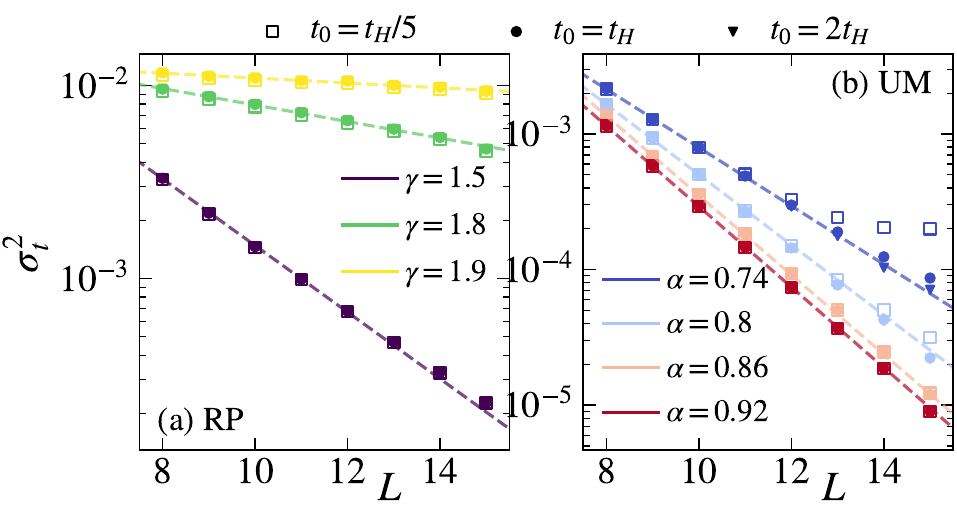}
    \caption
    {
        Variance of the temporal fluctuations $\sigma_t^2$, see Eq.~\eqref{eq:temporal}, versus the system size $L$, calculated from the time intervals with different initial times $t_0$.
        (a) RP model at different $\gamma$, and (b) UM model at different $\alpha$. 
        The dashed lines show fits the function $\propto\mathcal{D}^{-2/\eta_t}$ to the results for $t_0=t_H$.
    }\label{figA5}
\end{figure}

Finally, we study the role of the choice of time window to obtain the variances of temporal fluctuations $\sigma_t^2$ defined in Eqs.~\eqref{eq:temporal_state} and~\eqref{eq:temporal} of the main text.
We calculate the variances from the time interval $t\in[t_0,t_\infty]$, and here we focus on the role of the initial time $t_0$, while the final time is set to $t_\infty=10t_H$ as in the main text.

In Fig.~\ref{figA5}, we consider three different initial times: $t_0=t_H/5$, $t_0=t_H$ and $t_0=2t_H$.
We find that by choosing $t_0<t_H$, in the vicinity of the ergodicity breaking transition, $\sigma_t^2$ does not show a clear exponential decay with the system size $L$.
This effect can be attributed to the property that the system might not yet have reached a steady state after a quantum quench close to the ergodicity breaking transition.
Nonetheless, when choosing $t_0\geq t_H$, the exponential decay of $\sigma_t^2$ is restored, as observed in Fig.~\ref{figA5}(a) for the RP model and in Fig.~\ref{figA5}(b) for the UM model.
We note that the exponential decay is less pronounced in the RP model at large $\gamma$ due to the lack of equilibration at the ergodicity breaking transition. 
To mitigate the deviations near the transition and fixing a timescale that is not too large, we set $t_0=t_H$ in the main text.
\end{document}